\documentclass[times,twocolumn,final]{elsarticle}

\usepackage{medima}
\usepackage{framed,multirow}

\usepackage{amsmath,amssymb,amsfonts}
\usepackage{latexsym}
\usepackage{booktabs}
\usepackage{array}
\usepackage{caption}
\usepackage{tabularx}
\usepackage{graphicx}
\usepackage{url}
\usepackage{xcolor}

\usepackage{hyperref}
\newcolumntype{P}[1]{>{\centering}p{#1}}

\definecolor{newcolor}{rgb}{.8,.349,.1}

\journal{Medical Image Analysis}

\begin{document}

\verso{Kaiyan Li \textit{et~al.}}

\begin{frontmatter}

\title{O-PRESS: Boosting OCT axial resolution with Prior guidance, Recurrence, and Equivariant Self-Supervision}%

\author[1,2]{Kaiyan Li}
\author[3,4]{Jingyuan Yang}    
\author[1,2,7]{Wenxuan Liang}
\author[5]{Xingde Li}
\author[3,4]{Chenxi  Zhang}
\author[3,4]{Lulu Chen}
\author[3,4]{Chan Wu}
\author[3,4]{Xiao Zhang}
\author[3,4]{Zhiyan Xu}
\author[3,4]{Yuelin Wang}
\author[3,4]{Lihui Meng}
\author[1,2]{Yue Zhang}
\author[3,4]{Youxin \snm{Chen}\corref{cor1}}
\author[1,2,6]{S.Kevin \snm{Zhou}\corref{cor1}}
\cortext[cor1]{Corresponding author. 
  E-mail address: chenyx@pumc.cn (Youxin Chen), s.kevin.zhou@gmail.com (S.Kevin Zhou)}

\address[1]{School of Biomedical Engineering, University of Science and Technology of China, Hefei, 230026, China}
\address[2]{Center for Medical Imaging, Robotics, Analytic Computing \& Learning (MIRACLE), Suzhou Institute for Advanced Research, University of Science and Technology of China, Suzhou, 215123, China}
\address[3]{Department of Ophthalmology, Peking Union Medical College Hospital, Chinese Academy of Medical Sciences, Beijing, 100730,China}
\address[4]{Key Laboratory of Ocular Fundus Diseases, Chinese Academy of Medical Sciences,Beijing, 100730, China}
\address[5]{Department of Biomedical Engineering, Johns Hopkins Univesity, Baltimore, 21287, USA}
\address[6]{Key Lab of Intelligent Information Processing of Chinese Academy of Sciences (CAS), Institute of Computing Technology, CAS, Beijing,100190, China}
\address[7]{School of Physical Sciences, University of Science and Technology of China, Hefei, Anhui, 230026, China}


\begin{abstract}
Optical coherence tomography (OCT) is a noninvasive technology that enables real-time imaging of tissue microanatomies. The axial resolution of OCT is intrinsically constrained by the spectral bandwidth of the employed light source while maintaining a fixed center wavelength for a specific application. Physically extending this bandwidth faces strong limitations and requires a substantial cost. We present a novel computational approach, called as {\bf O-PRESS}, for boosting the axial resolution of {\bf O}CT with {\bf P}rior {\bf G}uidance, a {\bf R}ecurrent mechanism, and {\bf E}quivariant {\bf S}elf-{\bf S}upervision. 
Diverging from conventional superresolution methods that rely on physical models or data-driven techniques, our method seamlessly integrates OCT modeling and deep learning, enabling us to achieve real-time axial-resolution enhancement exclusively from measurements without a need for paired images. Our approach solves two primary tasks of resolution enhancement and noise reduction with one treatment. Both tasks are executed in a self-supervised manner, with equivariance imaging and free space priors guiding their respective processes. Experimental evaluations, encompassing both quantitative metrics and visual assessments, consistently verify the efficacy and superiority of our approach, which exhibits performance on par with fully supervised methods. Importantly, the robustness of our model is affirmed, showcasing its dual capability to enhance axial resolution while concurrently improving the signal-to-noise ratio. 
\end{abstract}

\end{frontmatter}


\section{Introduction}
\label{sec1}
Optical coherence tomography (OCT) is a volumetric imaging modality that allows non-invasive visualization of cross-sectional views of biological samples at a high axial resolution in real time~\citep{huang1991optical, fercher1996optical}. In ophthalmology, compared with color fundus images, which can only provide en-face information, OCT images provide cross-sectional information of all of the retina layers  \citep{swanson1993vivo, wojtkowski2002vivo, nassif2004vivo}. So segmentation and quantitative analysis of layers can be more accurate and convenient, which are very important for early detection, precise diagnosis and grading \citep{Farsiu2014,de2018clinically}. Insufficient resolution can significantly compromise the accuracy of segmenting the region of interest, leading to inaccurate diagnoses \citep{lee2017deep}. Therefore, achieving a high axial resolution is essential for reliable and precise diagnostic evaluations.

By taking advantage of the coherence gating of a light source, axial resolution of a micrometer ($\mu$m) scale is achieved. However, this also sets a limit on the axial resolution to the temporal coherence length, which is inversely proportional to the bandwidth of the light source \citep{izatt1996optical}. Traditionally, it has been believed that increasing the axial resolution of OCT images requires physically extending the spectral bandwidth of the system while maintaining a certain central wavelength \citep{drexler1999vivo,liu2011imaging}. However, either for spectral domain OCT (SD-OCT) or swept-source OCT (SS-OCT), this physical approach is limited by factors such as system complexity, available gain material, dispersion compensation, and spectral efficiency of the detector \citep{fercher2001numerical, wojtkowski2004ultrahigh, hu2007fourier,szkulmowski2016spectrometer, klein2017high}, which all affect the practical feasibility of expanding the physical bandwidth. Most importantly, an optical "window" centered at $\sim$1060 µm, the available spectrum bandwidth allowing just enough light (if following the safety guidelines of ANSI \citep{institute2007american}) transmitted to the fundus for obtaining high contrast images, theoretically limits the OCT axial resolution to $\sim$3.6 µm in retina \citep{hariri2009limiting}. Decreasing central wavelength is another option, however, scattering in tissue is much stronger at a shorter wavelength resulting in a shallower imaging depth \citep{povavzay2003enhanced, unterhuber2005vivo, povavzay2007three}. 

Given the existing dilemma and the rapid advancements in computer science, researchers are exploring alternative computational methods other than conventional deconvolution \citep{schmitt1997,Wang1999}, to further enhance axial resolution. Liu proposed an auto-regression-based spectral estimation technique that overcomes the limitations imposed by DFT \citep{Liu2015}. Various approaches, such as the iterative adaptive approach \citep{Wit2021}, alternating direction method of multipliers \citep{ling2020beyond}, and GPU-accelerated iterative method with regularization \citep{Wang2023}, have emerged and demonstrated significant improvements. These model-based optimization methods necessitate precise measurement of interference fringes and source spectrum. Nonetheless, in many clinical  applications that utilize  standard OCT settings, it is not feasible to expect practitioners to collect the source spectrum before each imaging operation. Furthermore, optimization has to be conducted for each frame reconstruction.  The iteration process, which consumes a considerable amount of time, remains a major drawback for real-time OCT imaging.

Deep learning has proven to be a powerful tool for enhancing the quality of OCT images \citep{halupka2018retinal, HuangOE2019, xu2020texture, Liang2020, Cao2020, lazaridis2021oct}. However, most of the existing research has primarily focused on addressing the issue of speckle reduction \citep{Ma2018, shi2019despecnet, huang2020noise, wang2021semi}, with only a limited number of papers investigating improvements in axial resolution \citep{HuangOE2019, Liang2020, Cao2020, Yuan2020, Zhang2021, Lee2023}. While these efforts show promise, the preferred approaches remain some major drawbacks as follows: 1) supervised learning utilized in these work typically relies on large-scale datasets containing paired low-resolution and high-resolution images. Unfortunately, acquiring such paired-datasets routinely in clinical practice is challenging; 2) current solutions to bypass the paired-requirement involve synthesizing low-resolution images from high-resolution measurements using techniques such as convolving a blur kernel with high-resolution images \citep{Liang2020}, or employing spectrum down-sampling or truncation \citep{Cao2020, Yuan2020, Zhang2021, Lee2023}. These methods hinder the generalizability of models to unknown measurement processes and  set a limit on the quality of model-generated images to the corresponding high-resolution images originally used as ground truth; 3) all published methods solely rely on neural network and ignore the physical relationship between the low and high resolution images, and thus can only generate high-resolution images with fixed degree of clarity. 

Reconstructing images from measurements is a fundamental challenge in imaging, and this holds true for OCT. The nature of OCT measurements inherently involves a band-pass process, which limits its ability to capture higher-frequency components that correspond to small structures in tissues. In this paper, we present a self-supervised model that addresses this limitation by learning a neural network to predict missing high-frequency contents for enhanced axial resolution of OCT. Our approach, called \textbf{O-PRESS}, has three distinct aspects: (i) it leverages the concept of \textit{equivariant self-supervision}to extend the spatial range of reconstruction space and enable the learning of high-frequency information;  (ii) It incorporates prior knowledge of OCT imaging to constrain the neural network within a tighter solution space in search of a unique and definitive solution, eliminating the need for paired supervision images and achieving deblurring effects using only a single measurement image in spatial domain without knowing the raw spectra; and (iii) it possesses a \textit{recurrent mechanism} for continuous improvement till convergence. Through the integration of the OCT imaging model with features extracted from the measured signals, our method demonstrates the significant potential for generating images with a higher axial resolution . Furthermore, our proposed approach employs a frame-to-frame processing strategy, in contrast to conventional methods that operate on a per A-line basis.

\begin{figure}[h]
\centering
\includegraphics[width=\columnwidth]{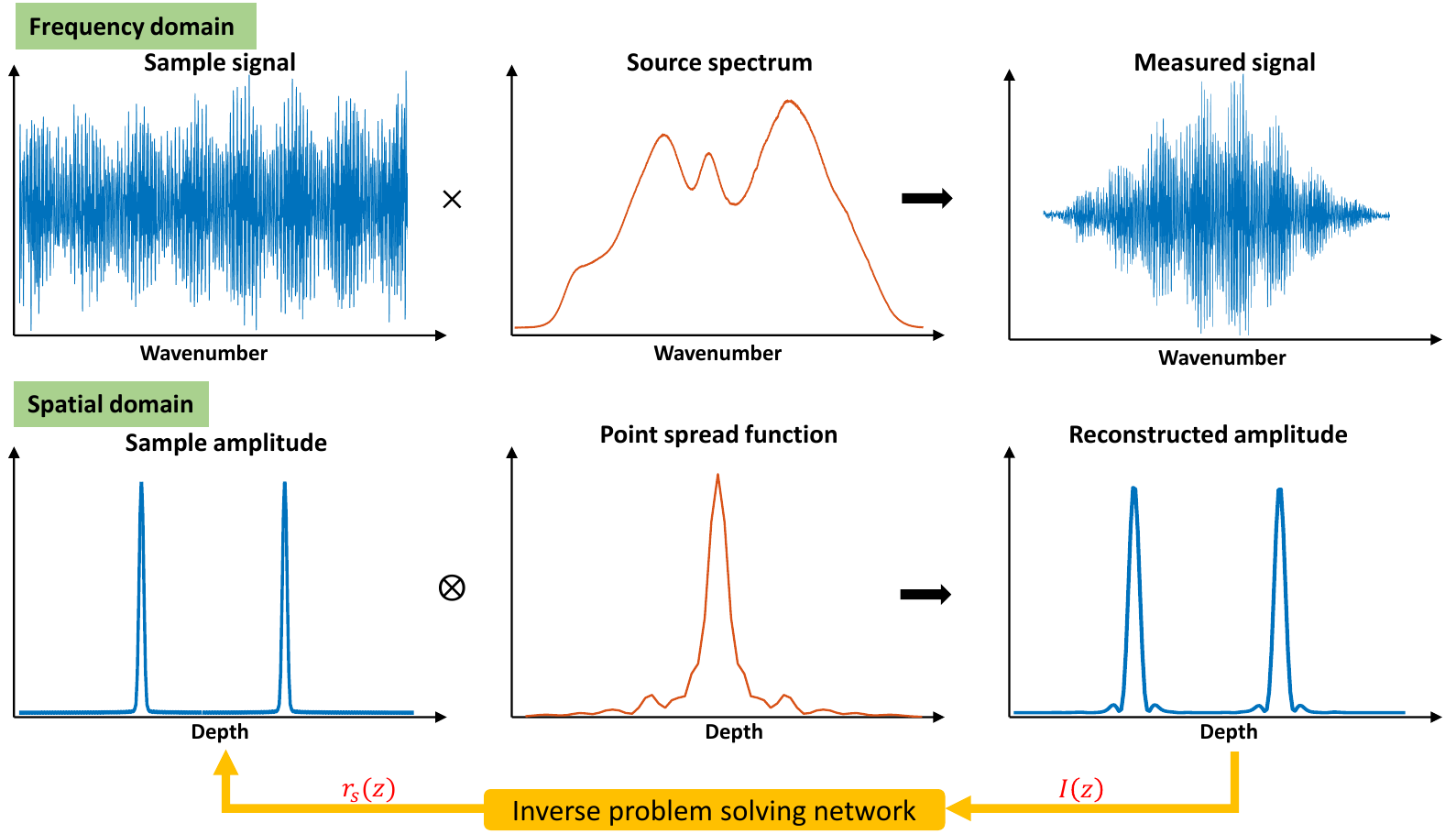}
\caption{
\begin{small}
    \textbf{Reconstruction architecture in frequency and spatial domains.} The sample signal represents the real signal of the sample. The source spectrum is the measured spectrum of the light source. The measured signal refers to the raw data in the frequency domain obtained from the Fourier domain OCT system. The reconstructed amplitude is the Discrete Fourier Transform (DFT) of the measured signal. The point spread function corresponds to the DFT of the source spectrum. The sample amplitude represents the real signal of the measured sample in the spatial domain. The primary objective of this study is to reconstruct the real profile $r_S(z)$ from the estimated amplitude $I(z)$. \end{small}}
\label{fig:oct}
\end{figure}

\section{The O-PRESS Method}
Below we elaborate the proposed O-PRESS method, which is a deep learning method that requires no paired supervision. Starting from a brief theory of OCT image reconstruction (Section \ref{sec.oct}), we then introduce the proposed loss functions that incorporate two imaging priors of equivariant imaging and free space in Section \ref{sec.prior}, which enables self-supervised learning, and the training and inference pipeline in Section \ref{sec.net}.  

\subsection{OCT image reconstruction} \label{sec.oct}
Fourier-domain OCT 2D images are reconstructed from the acquired spectra fringes by Fourier-transform,
\begin{equation}
    I(z) = \mathrm{IDFT}\{I(k)\};~~ I(k)=2S(k) r_R \mathrm{\sum_{n=1}^{N}} r_S(z_n) \mathrm{cos} (2kz_n) \delta_n,
\end{equation}
where $I(k)\in C^M$ is the measured signal, $S(k)\in C^M$ is the power spectral density, the reference reflectivity, $r_R$, is typically set to 1, and $r_S(z) \in C^N$ characterizes the sample reflectivity. When $S(k)$ has infinite bandwidth and the granularity of particles in the object space $\delta_n$ matches the digital resolution $\delta_m$ in the image space, $r_S(z)$ can be accurately estimated by performing the inverse discrete Fourier transform(IDFT) on the measured signal $I(z)=\mathrm{IDFT}\{I(k)\}$. However, two limiting factors impact the accuracy of this estimation. 

Firstly, the bandwidth of $S(k)$ is limited and the full width at half maximum (FWHM) of its Fourier transform, sets the upper limit of axial resolution. As depicted in Fig.~\ref{fig:oct}, in the measurement domain of FD-OCT, i.e. the frequency domain, the sample spectrum is shaped by the spectrum of light source. After Fourier transform, the reconstructed signal in the spatial domain is degraded by the point spread function (PSF) of the imaging system. Note that the PSF of an OCT scanner is a composite characteristic resulting from the combined effect of the light source and imaging optics. The Fourier transform of power spectral density, denoted as $h(z)$, dominates the PSF.

Secondly, imaging is a measurement process of mapping signals from a higher-dimensional object space (dimension \textit{N}) to a lower-dimensional image space (dimension \textit{M}). The particle granularity $\delta_n$ in the object space is much finer than the digital resolution $\delta_m$ in the image space, with $N>M$. Consequently, reconstruction of high-dimensional object signal from lower-dimensional measurement signal is an ill-posed inversion problem and hence challenging. 

As a result, the reconstructed A-line signal can be approximated as the convolution of $h(z) \in C^M$ and the actual signal $r_S(z) \in C^N$, plus the sum of sampling error and shot noise  $\epsilon$:
\begin{equation}
    I(z) \approx h(z)\ast r_S(z)+\epsilon. \label{eq:imaging}
\end{equation}
It is evident from Eq. (\ref{eq:imaging}) that reconstructing a high-resolution image $r_S(z)$ from a solitary low-resolution input $I(z)$ is inherently intricate as it involves both \emph{deblurring}, that is, removing the effect of the convolution kernel $h$,  and \emph{denoising}, that is, removing the effect of noise $\epsilon$.

\subsection{Equivariant imaging prior and free space prior} \label{sec.prior}
While supervised deep learning method is widely studied for OCT reconstruction, it has an inherent disadvantage, that is, it requires paired low-resolution images and high-resolution images. This requirement is practically different to fulfill. In this paper, we introduce a novel self-supervised deep learning based approach to enhance axial resolution in the spatial domain, which is specifically tailored for situations where only OCT intensity images are available. 

In self-supervised learning, the key is to define surrogate supervisory signals, which are often designed by utilizing the inherent structures or patterns within the data~\citep{shurrab2022self, huang2023self, chen2019self, taleb20203d, azizi2021big, bai2019self}. In this work, we propose to use equivariant imaging prior and free space prior, both customized for OCT.


\subsubsection*{Equivariant imaging (EI) prior} 
We employ the equivariant imaging strategy proposed by~\citep{Chen2021} to estimate a unique inverse mapping  $\Phi$.  Specifically, to accomplish the deblurring task in OCT, we leverage the \emph{shift invariance property} of OCT imaging, assuming that PSF is constant within the effective imaging depth: shifting A-lines both in the fast scanning axis and along the imaging depth before reconstruction is equivariant to shifting A-lines after reconstruction:
\begin{equation}
    \Phi \circ h \ast (T_g r_S) = T_g \Phi \circ (h \ast r_S),
\end{equation}
where $T_g$ defines shift transformations. It is worth emphasizing that it is the composition of $\Phi \circ h$ is equivariant to shift transformation in OCT imaging. Here, one observation of the measurement corresponds to one shift transformation. In order to recover the signal  $r_S(z) \in C^N$ from partial measurements $I(z)\in C^M$, a necessary condition is that the range space spanned by multiple observations of the measurement covers the full range space $\mathbb{R}^N$:

\begin{equation}
\begin{bmatrix}
     h \ast (T_1 r_S)  \\
     \vdots \\
     h \ast (T_g r_S) 
\end{bmatrix}
\in \mathbb{R}^N
\end{equation}
This condition requires $gM \geqslant N$, in order to guarantee that a unique inverse mapping can be learnt from the concatenated observations.

The equivariant imaging prior constraint allows the expansion of the spatial range of reconstruction space by incorporating different observations and enables the network to learn high frequency information beyond the original measurement domain. In other words, it has a \textbf{deblurring effect}.

\subsubsection*{Free space (FS) prior} Equivariant learning provides a unique solution to under-defined inverse problems when the forward model is known, such as when the PSF is determined. It enables the network to recover high-resolution details from a single low-resolution image by generating high frequencies. However, when the PSF is not precisely known and only its estimation is available, equivariance learning introduces additional noise, leading to significant artifacts (as shown in the EI image in Fig. \ref{fig:fig2}e). 
In light of this challenge, we have developed an effective approach to tackle the \textbf{denoising} task by utilizing prior knowledge about the imaging properties of OCT. 

In a normal eye, the aqueous humor above the retina surface is transparent and exhibits clear and low signal intensity in OCT images. We segment this region out (illustrated as the Mask in Fig. \ref{fig:fig2}e) and set it to zero during training as denoising guidance. We call this as \textit{free space prior}. Similar strategies can be employed for different OCT applications other than eyes, as one can always set a low signal gap between the tissue surface and the zero-delay position determined by the reference arm.

\subsubsection*{Method for generating free space masks}
For correctly generate the free space mask, we select only normal retina images for training as their free space regions are clear and of low-signal, and thus can be set to zeros. And also, it is easier to segment them out if no bleeding or opacity in it.  In principle, all segmentation methods should work in this task. In this study, we use a simple gradient-based thresholding approach as the tissue surfaces in OCT images were mostly apparent. 
\begin{equation}
    M = \pi[~ | \nabla (BLF(I))|  > \tau~],
\end{equation}
where $M$ is the segmentation mask, $\pi[.]$ is an indication function, $\nabla$ is gradient operator, $BLF$ is a bilateral filter that preserves the edge during smoothing, and $\tau$ is a threshold (we empirically set it to 3).
All the generated masks are manually checked to ensure that all training images are correctly segmented. Examples of the generated masks are shown in Supplementary Fig.\ref{fig:mask}.

We also try the Segment Anything Model (SAM)~\citep{kirillov2023segment} for this task. It turns out that our simple gradient-based thresholding method outperforms the complex SAM method.

\begin{figure*}[t]
\centering
\includegraphics[width=\textwidth]{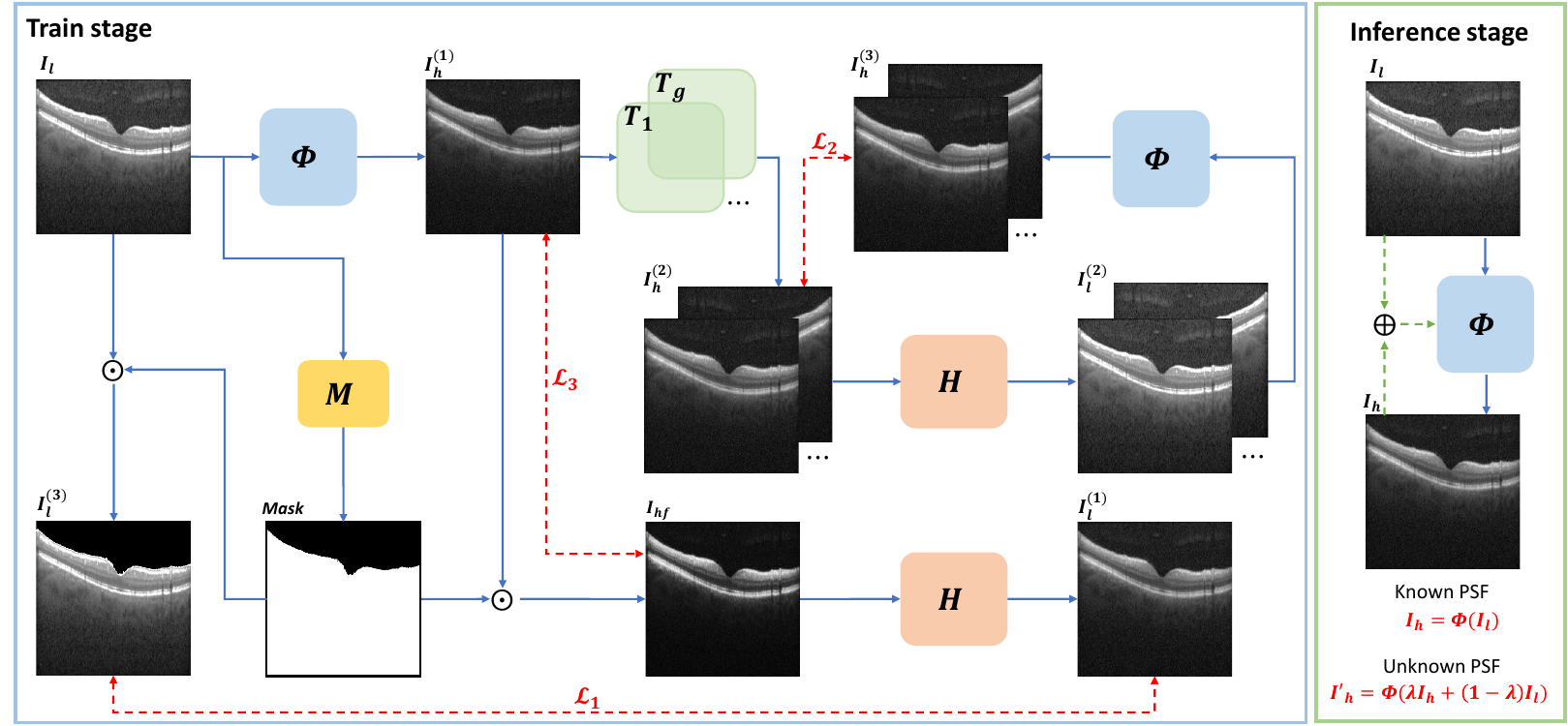}
\caption{
\begin{small}
    \textbf{Pipeline of the proposed O-PRESS Framework.} The symbol "$\odot$" denotes pixel-wise multiplication. "$\mathit{\Phi}$" signifies the high-resolution reconstruction model. "$\mathcal{M}$" denotes the mask generation (aka segmentation) module. "$T_g$" denotes the shift transformation along the fast scanning axis.  "$H$" simulates the OCT imaging process, equivalent to the convolution of the high-resolution signal with a PSF.
\end{small}}
\label{fig:fig1}
\end{figure*}
\subsection{Neural network learning and inference} \label{sec.net}
Fig.~\ref{fig:fig1} depicts the pipeline of our method during training and inference. In the pre-processing step, low-resolution images are interpolated in the axial direction to achieve a digital resolution at least twice as high as the optical resolution, resulting in $I_l$. The neural network (denoted as $\Phi$), based on the ResU-Net architecture~\citep{Chen2021}, takes $I_l$ as inputs and generates high resolution images $I^{(1)}_h$, while a segmentation module (denoted as $M$) also takes $I_l$ as inputs for mask generation.
\begin{equation}
    I^{(1)}_h = \Phi(I_l).
\end{equation}

To train the network to achieve the super-resolution goal, we employ three loss functions together. The \textit{first loss} function is the measurement consistency (MC) loss, denoted as $L_1$. It measures the difference between two sources: $I^{(1)}_l$ and $I^{(3)}_l$. $I^{(1)}_l$ is generated as follows: The foreground of the generated $I^{(1)}_h$ is imaged by the OCT system (module $H$) and hence degraded to low-resolution $I^{(1)}_l$. $I^{(3)}_l$ is the foreground of the input $I_l$. 
\begin{equation}
L_1 = \|I^{(3)}_l-I^{(1)}_l\|_2=\| (I_l -  H( \Phi(I_l)) \odot M \|_2.
\end{equation}
The \textit{second loss} function is the equivariance loss, denoted as $L_2$. $I^{(1)}_h$ is shift transformed several times, and the resulting images are concatenated to form $I^{(2)}_h$. $I^{(2)}_h$ is imaged by the OCT system (module $H$) and degraded to low-resolution images $I^{(2)}_l$. After applying the network, new high-resolution images $I^{(3)}_h$ are obtained. $L_2$ is calculated as the difference between $I^{(2)}_h$ and $I^{(3)}_h$. 
\begin{equation}
L_2 = \|I^{(2)}_h-I^{(3)}_h\|_2=\|  T_g \Phi(I_l)  - \Phi(H(T_g \Phi(I_l)))\|_2. 
\end{equation}
The \textit{third loss} function is the free space loss, denoted as $L_3$, which is constructed as the difference between $I^{(1)}_h$ and $I_{hf}$ using the free space prior. Minimizing $L_3$ enforces the region above the retina in normal eyes to exhibit clear and low signal intensity. 
\begin{equation}
L_3=\|I^{(1)}_h-I_{hf}\|_2=\|\Phi(I_l) \odot (1-M) \|_2.
\end{equation}
The overall loss function is a combination of these three objectives:
\begin{equation}
    L = \lambda_1 L_1 + \lambda_2 L_2 + \lambda_3 L_3,
\end{equation}
where $\lambda_1$, $\lambda_2$, and $\lambda_3$ represent the weights of MC, EI loss, and FS loss, respectively. Their values for training the models used in experiments below can be found in Supplementary Table \ref{tab2}. 

\subsubsection*{Training}
There are two distinct training approaches based on the situation of the PSF. In cases where the PSF is known, it can be directly applied in the computational imaging process by convolving with high-resolution images (module \textit{H} in Fig. \ref{fig:fig1}) to produce low-resolution counterparts. In this scenario, the FS guidance can be optionally employed, as the EI guidance alone typically yields satisfactory results. However, when the PSF is unknown, it needs to be estimated based on system parameters. We initiate the PSF estimation by utilizing a Gaussian function, with the standard deviation determined by the axial resolution of the system. The network then compensates for side-lobe effects. In this situation, the FS guidance becomes essential. It collaborates with the EI guidance to generate enhanced images with improved resolution and signal-to-noise ratio (SNR). For the purpose of improving the generalization capability, a smaller standard deviation for the PSF is chosen. This enables the trained model to be applied on datasets with an originally higher resolution for further enhancement. In such a case, a recurrent strategy is employed during the inference stage to progressively increase the resolution, as depicted in Fig. \ref{fig:fig1}. 
\begin{equation}
I_h = \Phi(I_l); ~I'_h = \Phi(\lambda I_h + (1-\lambda) I_l),
\end{equation}
where $I'_h$ is the recurrent output and $\lambda$ is a linear blending coefficient.
Hyperparameters used in different training strategies are summarized in Supplementary Table \ref{tab6}. The Adam optimizer is used to train the established network.

\subsubsection*{Inference}
During the inference stage, two imaging scenarios are considered: 
1) When the PSF is known, a low-resolution image $I_l$ is fed into the neural network $\Phi$, which predicts a high-resolution image $I_h=\Phi(I_l)$;
2) In cases where the PSF is unknown, in order to achieve a unified one-model-for-all approach, a recurrent mechanism is employed. As shown in the green box of Fig.\ref{fig:fig1}, the high-resolution image is recurrently reconstructed as $I'_h= \Phi(\lambda I_l + (1-\lambda) I_h)$. This recurrence enables our method to adapt to enhancing images at different levels of clarity. Supplementary Fig.~\ref{fig:recurrent} demonstrates such a continuous improvement in resolution until convergence.

\section{Results and Discussions}
\subsection{Datasets}
The training data for our model is sourced from a publicly available dataset \citep{Kermany2018} from which we select 600 retina images. Out of these, 450 images of normal eye are utilized for training the model, while 100 images are set aside for validation purposes. The remaining 50 images are specifically reserved for testing the model's performance. Five-fold validation is performed to makeup the insufficient data size and avoid over-fitting. The imaging system employed for capturing the retina images is the Spectralis OCT from Heidelberg. The configuration details of this OCT system, along with the corresponding configurations of systems used for collecting validation data at the purpose to assess the model's generalization, are presented in Supplementary Table \ref{tab5}.

To evaluate the generalization capability of our method across different OCT systems and resolutions, we collect additional two OCT datasets (Datasets 2 and 3 listed in Supplementary Table \ref{tab5}) of human retina using the Topcon Triton system and SVision DREAM OCT system from Peking Union Medical College Hospital (PUMCH). Furthermore, to assess the robustness of our model, we acquire OCT images of abnormal human retina with only 2 frame averaging using the SVision DREAM OCT system at PUMCH (Dataset 4 listed in  Supplementary Table \ref{tab5}). In addition, we conduct a validation using esophagus mucosa samples of swines (Dataset 5 listed in Supplementary Table \ref{tab5}) obtained from the BIT lab at Johns Hopkins University (JHU) \citep{Li2019} to evaluate the generalization ability of our method across different tissue types. Detailed system description, animal experiments and the followed protocols for collecting OCT images of esophagus mucosa are described in the reference \citep{Li2019}.

The study is approved by the Institutional Review Board of Peking Union Medical College Hospital (No. HS-2174) and ethics committees. Written informed consent is obtained from the participants. The study follows the tenets of the Declaration of Helsinki principles. Common types of retina disease are included in the study for analysis. 
Dataset 2, obtained from 10 patients, has a moderate resolution and high SNR. Dataset 3, acquired from 6 patients, has a high resolution and high SNR. Dataset 4, gathered from 4 patients, has a high resolution but a low SNR.

\subsection{Evaluation and metrics}
To demonstrate the accuracy and efficiency of our model, with a particular focus on the efficacy of EI and FS priors, we carry out a series of experiments on both simulated and real OCT data. Furthermore, we conduct comparative analyses against Richardson-Lucy (RL) deconvolution and supervised learning to underscore the advantages of our proposed approach. 

Evaluating the quality and realism of the reconstructed images from the network is a challenging and ongoing problem. We assess it using both quantitative and qualitative measures.

\subsubsection*{Quantitative assessment} 
When evaluating the performance of our method on a dataset that includes paired ground truth data, we utilize the PSNR and SSIM as assessment indices to quantitatively measure the image fidelity. In cases where the dataset only consists of low-resolution images without ground truth, we employ the following evaluation metrics~\citep{Ma2018}.

\underline{Edge preservation index (EPI)}. It reflects the ability of the network to preserve the edge details after processing. Because the edges or layers of OCT images mainly exist in the axial direction, EPI is defined as:
\begin{equation}
    EPI=\frac{\sum_i \sum_j \left |{I_n(i+1, j)}-I_n(i, j) \right |}{\sum_i \sum_j \left |{I(i+1, j)}-I(i, j) \right |},
\end{equation}
where $I_n$  represents the network generated image, $I$  is the original low-resolution  image, and $i$ and $j$ denote the \textit{i}-th row and \textit{j}-th column of the image, respectively.

\underline{Equivalent number of looks (ENL)}. This index is used to measure the smoothness of the homogeneous region in the processed image, which is widely employed in evaluation the performance of speckle reduction in OCT image. It quantifies the degree of noise suppression achieved by the network. It is calculated over the background ROI of images in the following manner:
\begin{equation}
    ENL=\frac{\mu^2_b}{\sigma^2_b}
\end{equation}
where $\mu_b$ and $\sigma_b$ represent the mean and standard deviation of selected background ROI in each image, respectively. 

\underline{Signal-to-noise ratio (SNR)} and \underline{contrast-to-noise (CNR)} serve as indicators of the noise level in relation to the signal level and the contrast between the signal in the ROI and the noisy background. In the SNR measurement for OCT, noise floor is obtained by blocking the sample arm during imaging to exclude sample signal~\citep{yun2003high}.  In this study, we consider the background region above the tissue surface as the noise area, as it does not contain any relevant signal of interest~\citep{de2003improved, agrawal2017methods,baumann2019signal, Ma2018, Wang2023}.
\begin{equation}
    SNR=10\mathrm{log}10(\frac{\mu_s^2}{\sigma_b^2});~~    CNR=10\mathrm{log}10(\frac{\mu_s-\mu_b}{\sqrt{\sigma_s^2+\sigma_b^2}}).
\end{equation}

\subsubsection*{Qualitative assessment}
In our experiment, a total of ten human experts participated in the evaluation of neural network-enhanced images. The panel includes eight ophthalmologists, consisting of two students, two ophthalmology residents, two intermediate ophthalmologists, and two senior ophthalmologists. Additionally, one senior OCT expert and one senior optical imaging expert are also involved. These experts are not involved in the training process and are unaware of the trained models and the model-generated images beforehand. The evaluation is conducted in a ``paired" manner, where a generated retina image is presented alongside its original input image, with the order of the images randomized. 
Optical experts are asked to choose the higher quality image based on criteria such as resolution, contrast, layer clarity, and SNR. Ophthalmologists, on the other hand, are assigned the responsibility of assessing the images in terms of resolution, layer clarity, and overall image quality that is valuable for diagnosis.
Following each evaluation, the obtained results are statistically analyzed to calculate ``quality scores" for each evaluation criterion. A quality score of 1 indicates that the generated images exhibit superior image quality compared to the input. Conversely, a score of 0 indicates that the image quality of the generated image is lower than that of the input. A score of 0.5 suggests that the output and the input exhibit similar levels of image quality.

\begin{figure*}[ht!]
\centering
\includegraphics[width=\textwidth]{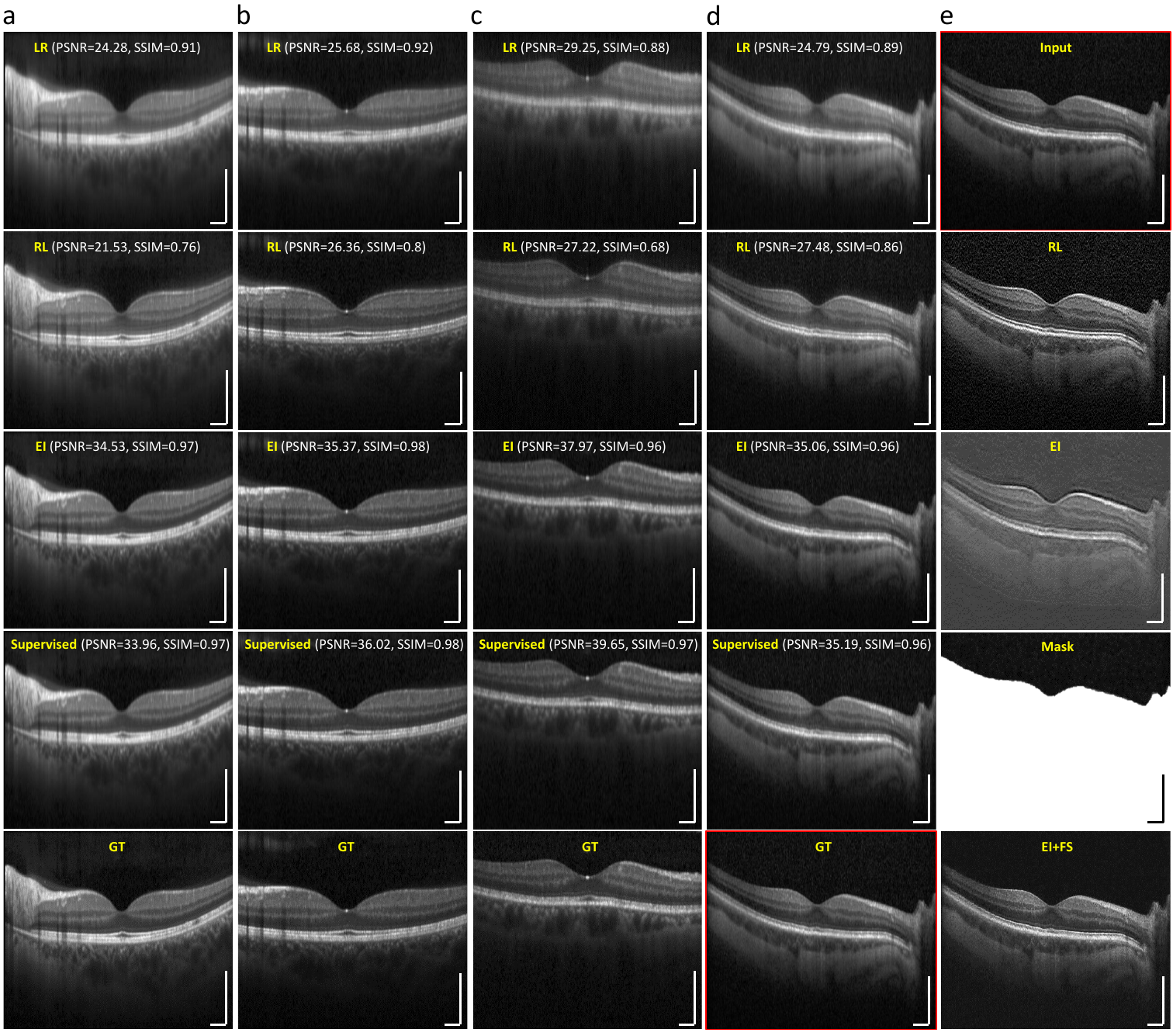}
\caption{
\begin{small}
 \textbf{Comparison between RL deconvolution, supervised-learning and our approach.} \textbf{a-d}, super-resolution reconstruction results using different methods with a known real PSF. LR: low resolution images, RL: reconstructed images using Richardson-Lucy deconvolution with a kernel estimated by a Gaussian function. The full-with-half-maximum (FWHM) of the Gaussian function is the same with the known real PSF that blurred the ground truth (GT). EI: reconstruction images of our self-supervised model guided by EI prior without GT pairs. Supervised: images of a supervised model trained with GT pairs. The corresponding PSNR and SSIM values are indicated in parentheses. \textbf{e}, super-resolution reconstruction results without knowing the PSF and GT. The input is the same image as GT in \textbf{d}. ES+FS: the reconstructed image of our self-supervised model guided by both EI and FS priors. Scale bar: 0.5 mm.
\end{small} }
\label{fig:fig2}
\end{figure*}

\begin{table*}[!h]
    \centering
    \footnotesize
    \renewcommand{\arraystretch}{1.5}
    \setlength{\tabcolsep}{5pt}
    \caption{Evaluation metrics for different methods on simulated low-resolution retina images}
    \vspace{0pt}
    \captionsetup{skip=3pt} 
    \scalebox{1.05}{
    \begin{tabular}{lrrrrrrrr}
    \hline
    &LR&RL&SUP&EI\_1-fold&EI\_2-fold&EI\_3-fold&EI\_4-fold&EI\_5-fold \\
    \hline
    PSNR(dB) & 22.411$\pm$1.387 & 28.031$\pm$3.460 & 35.375$\pm$2.059 & 34.927$\pm$2.061 & 36.907$\pm$1.988 & 35.217$\pm$2.033 & 33.915$\pm$2.125 & 35.905$\pm$1.993 \\
    SSIM & 0.894$\pm$0.059 & 0.837$\pm$0.037 & 0.967$\pm$0.011 & 0.959$\pm$0.013 & 0.978$\pm$0.011 & 0.966$\pm$0.013 & 0.951$\pm$0.015 & 0.972$\pm$0.010 \\
    \hline
    \end{tabular}} 
    \label{tab1}
\end{table*}

\subsection{Results}
\subsubsection*{Overall predictive performance and comparative analysis}
To showcase the effectiveness of the EI prior, we train a model ($model_{ei}$) on the training set of dataset 1 (Supplementary Table \ref{tab5}) and evaluate its performance on the test set. The model takes low resolution images (LR) synthesized by convolving the ground truth (GT) with a real PSF ( in Supplementary Fig.~\ref{fig:psf}a)  along the imaging depth as input. By comparing the reconstructed images with the GT images, we assess and compare the reconstruction results of several methods. The results reveal that RL deconvolution performs admirably when the PSF resembles a Gaussian shape. However, in real-world scenarios with the PSF deviated from a pure Gaussian function, such as in this work with asymmetrical side lobes, the performance of RL deconvolution falls short. As depicted in Fig.~\ref{fig:fig2}a-d, the images reconstructed by RL deconvolution method show good resolution enhancement, but the restoration of speckle intensity and background is poor, leading to limited improvements in PSNR (peak signal-to-noise ratio) and a decrease in SSIM (structural similarity index measure). On the other hand, our self-supervised EI method achieves significant improvements in resolution while preserving the speckle intensity, which is nearly comparable to the results obtained through supervised approaches, both visually and based on assessment metrics. This demonstrates the successful learning of missing high frequency components. A statistical summary of PSNR and SSIM calculated from the test set of dataset 1 in Table~\ref{tab1} further underlines the effectiveness of our self-supervised method. Additionally, we conduct a 5-fold cross-validation to showcase the stability and generalizability of our model. The corresponding PSNR and SSIM of each fold are also displayed in Table~\ref{tab1}.

However, in situations where the PSF is unknown, which is often the case in real-world applications, the FS prior becomes valuable. We train a model named $model_s$ on the same dataset as above, but with an estimated PSF as a Gaussian function (supplementary Fig.~ \ref{fig:psf}b, the same one as used in RL deconvolution method). The FWHM bandwidth of the Gaussian function is set equal to the axial resolution specified in the system's manual. In this case, the training set is directly used as input for the model without undergoing any blurring process. By incorporating both the EI and FS guidance, the trained model successfully improves the resolution of the input, as shown in the image labeled EI+FS in Fig.~\ref{fig:fig2}e. Notably, this image also maintains satisfactory speckle intensity and noise level, surpassing the results obtained by a model trained solely with EI guidance and the RL deconvolution method. It is important to highlight that the input image in Fig.~\ref{fig:fig2}e is identical to the GT image in  Fig.~\ref{fig:fig2}d, and thus the resolution of the EI+FS image is even higher, demonstrating the effective improvements without relying on ground truth information.  Since the RL deconvolution method yields no good results in both cases, we do not use or compare it in the subsequent experiments.

\begin{figure*}[ht]
\centering
\includegraphics[width=\textwidth]{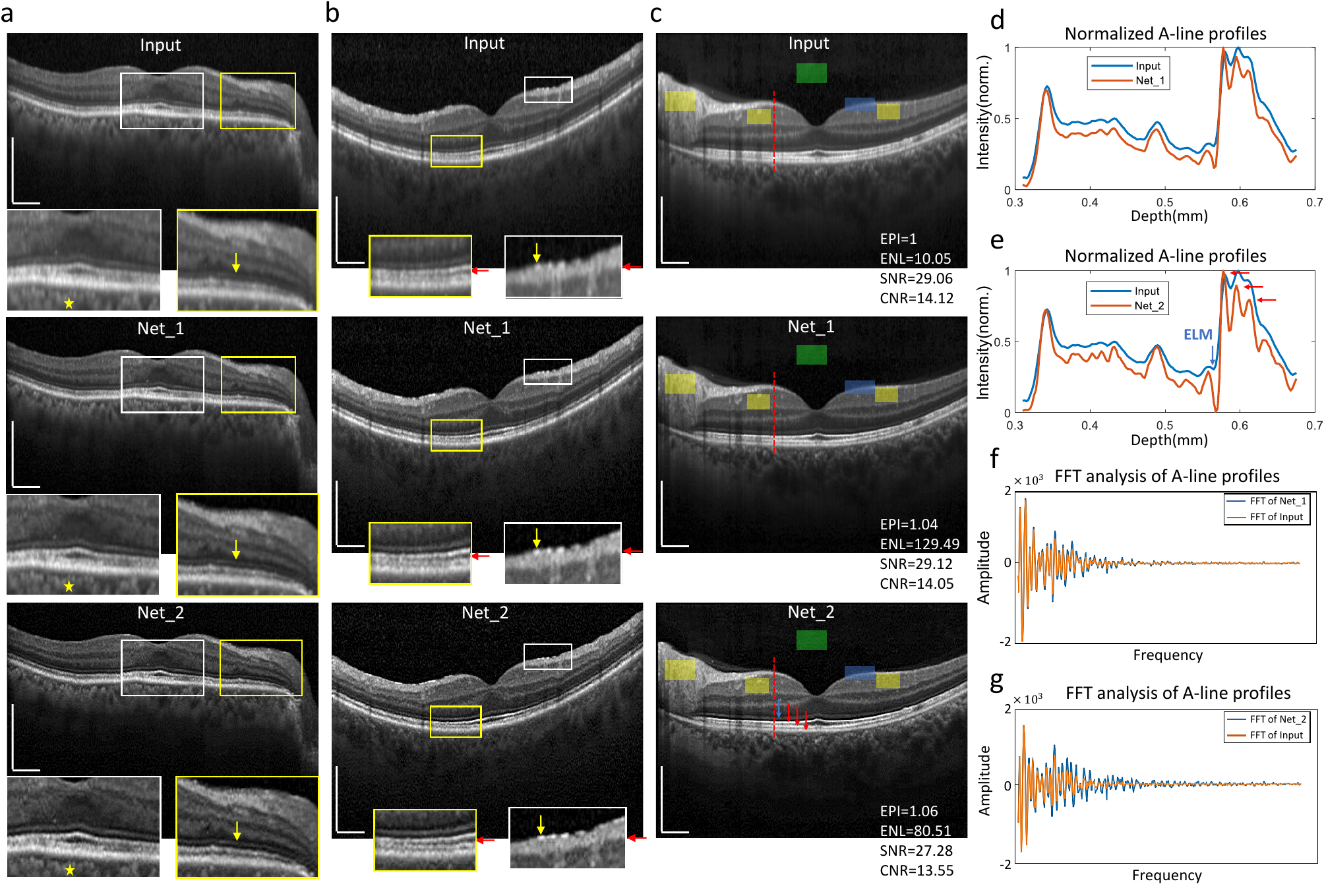}
\caption{
\begin{small}
  \textbf{Recurrent inference and resolution quantification.}  \textbf{a-c}, Input: retina OCT images from dataset 1 (Spectralist in Supplementary Table \ref{tab5}), Net\_1 = $model_r$(Input), Net\_2 = $model_r$(0.3 Input+0.7 Net\_1), insertions are 2 times zoom-in figures, green patches in \textbf{c} are regions selected as background, blue patches are regions selected for EPI calculation and yellow patches are selected for CNR calculation. \textbf{d-e}, normalized intensity of a cropped rang of A-lines labeled by red dashed lines in c. \textbf{f-g}, FFT analysis of the whole A-lines at the same horizontal locations as in d and e. Scale bar: 0.5 mm.
\end{small} }
\label{fig:fig3}
\end{figure*}

Fig.~\ref{fig:fig3} illustrates the gradual refinement of input images through the utilization of a recurrent inference strategy. The training process follows the same procedure as $model_s$, except for the use of a narrower PSF (Supplementary Fig.~\ref{fig:psf}c).  The trained model is named as $model_r$. The zoom-in images in Fig.~\ref{fig:fig3} clearly illustrate that Net\_1 images exhibit a higher resolution compared to their inputs, while Net\_2 images achieve an even higher resolution than Net\_1. In  Fig.~ \ref{fig:fig3}a, yellow arrows within yellow boxes indicate the external limiting membrane (ELM), which is resolved with the highest clarity in Net\_2. Net\_1 also shows improved visibility compared to the input. For the choroid, denoted by yellow stars in white boxes, Net\_2 achieves the highest level of clarity, and Net\_1 also shows significant improvement compared to the input. In  Fig.~\ref{fig:fig3}b, within the zoomed-in white boxes, a small hyper-reflective foci indicated by a yellow arrow appears blurry in the input image but becomes much clearer in Net\_1. In Net\_2, not only is the small structure easily isolated from the nerve fiber layer below, but the boundaries of other retinal layers, indicated by red arrows, are also much clearer and sharper. These observations collectively demonstrate the superior resolution achieved by Net\_2.

To quantitatively evaluate the resolution enhancement capability of our model, we calculate and provide EPI, ENL, SNR, and CNR values in the bottom right corners of the images in Fig.~ \ref{fig:fig3}c. Upon analyzing these values,  it becomes evident that Net\_1 achieves the highest ENL and SNR, whereas Net\_2 demonstrates the superior EPI. Taking into account both the detailed reconstruction performance and the evaluation metrics, we conclude that the recurrence technique facilitates continuous resolution improvement but also amplifies the noise level. Therefore, a balance must be struck between resolution and SNR. Figs.~\ref{fig:fig3}d-e depict the averaged A-line intensities of five adjacent A-lines of the input and `Net' images in Fig.~\ref{fig:fig3}c. By examining the intensity distribution, we are able to evaluate resolution by identifying and quantifying layer boundaries. The ELM and the three layers below it (red arrows in Figs.~ \ref{fig:fig3}c and \ref{fig:fig3}e) are most clearly discernible in Net\_2, once again demonstrating the effectiveness of our method. Figs.~\ref{fig:fig3}f-g present the results of Fourier transform analysis. Net\_1 exhibits more high frequencies associated with small structures and details compared to the input, while Net\_2 contains even more high frequencies related not only to details but also to additional noise.

\begin{figure}[ht!]
\centering
\includegraphics[width=\columnwidth]{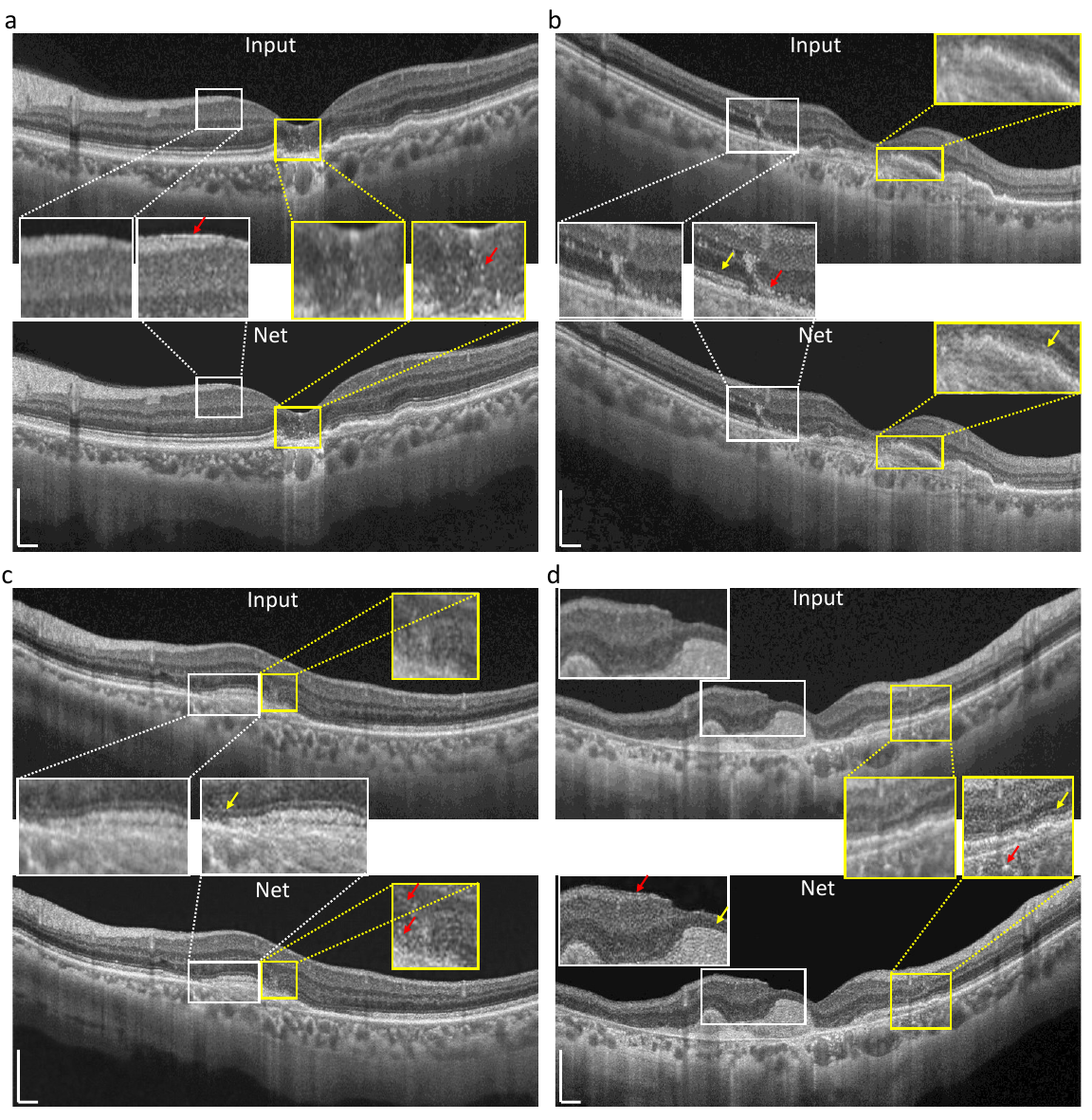}
\caption{
\begin{small}
  \textbf{Validation of the generalization capability of a model ($model_s$) trained on dataset 1 (Spectralist retina images) with a broader PSF on dataset 2 (Topcon retina images).} \textbf{a}, an OCT cross-sectional image of outer retinal atrophy in the fovea. The red arrow within the white box points to the inner limiting membrane. In the yellow box, the red arrow indicates intra-retina hyper-reflective foci in the Macular Fovea. \textbf{b}, an OCT cross-sectional image of neovascular pigment epithelial detachment with retinal exudation. Within the white box, the yellow arrow displays the external limiting membrane, while the red arrow indicates hyper-reflection in the disorganized outer layer of retina, which represents the residual ellipsoid zone. Within the yellow box, the yellow arrow indicates the layer of retinal pigment epithelium. \textbf{c}, an OCT cross-sectional image of choroidal neovascularization. Within the white box, the yellow arrow indicates the distinct external limiting membrane, while within the yellow box, red arrows highlight clearer, small retinal changes that may indicate inflammatory cells or exudation. \textbf{d}, an OCT cross-sectional image of scar formation of choroidal neovascularization with outer retinal atrophy. Within the white box, the red arrow indicates inner limiting membrane, while the yellow arrow indicates disrupted external limiting membrane. Within the yellow box, the yellow arrow indicates the external limiting membrane, while the red arrow indicates a hyper-reflective foci within the choroid region. Scale bar: 0.5 mm.
\end{small} }
\label{fig:lesion}
\end{figure}

\subsubsection*{Validation of cross-system generalization capability on high SNR retina images}
Before validation, adjustments are made to the intensity and digital resolution of the input images to ensure their similarity to the training data, thereby guaranteeing consistent and optimal performance. We present two examples to showcase the versatility of our proposed method.    

\begin{figure*}[h]
\centering
\includegraphics[width=\textwidth]{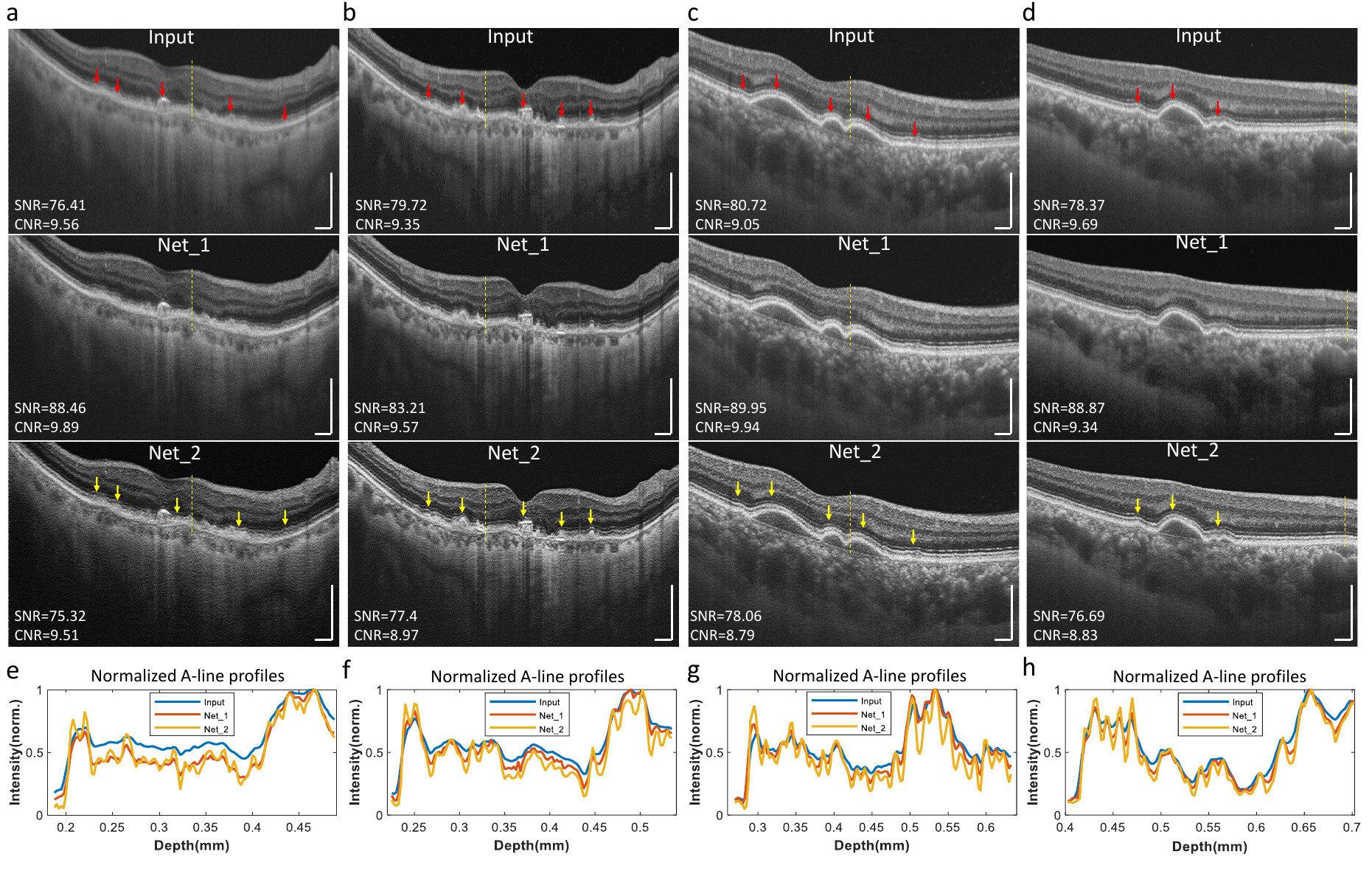}
\caption{
\begin{small}
   \textbf{Validation of the recurrent inference capability by recurrently applying a model trained on dataset 1 with a narrower PSF to dataset 2.} \textbf{a-d}, four examples demonstrating the improvement in resolution achieved through this recurrent process. Net\_1 = $model_r$(Input), Net\_2 = $model_r$(0.3 Input+0.7 Net\_1). The red arrows highlight pathological changes: reticular pseudodrusen in \textbf{a}, drusen and reticular pseudodrusen in \textbf{b}, and drusen in \textbf{c-d}. The yellow arrows indicate the resolution-enhanced ELM.
\textbf{e-h}, A-line profiles extracted from Input, Net\_1 and Net\_2. The yellow dashed lines in panels \textbf{a-d} mark the specified range of the plotted A-lines. Scale bar: 0.5 mm.
\end{small} }
\label{fig:fig4}
\end{figure*}

Firstly, $model_s$ is utilized for the direct reconstruction of high-resolution images (`Net') from low-resolution images (`Input'). Fig.~\ref{fig:lesion} illustrates four representative pairs of low-resolution retina OCT images alongside their corresponding high-resolution reconstructions. A comparison between the images of `Input' and `Net' reveals that the boundaries of retinal structures and lesions in the `Net' images are more distinct. This improvement in resolution and contrast enhances the ability of OCT to discern image details and small changes without introducing any observable distortions. Without a doubt, this makes it easier and more reliable for doctors to make accurate diagnoses. Furthermore, it is important to acknowledge that the input images, acquired by averaging 32 frames, exhibit high levels of SNR and CNR. The pursuit of improving resolution in the `Net' images, accomplished by generating high-frequency components through the use of the EI and FS priors, effectively further diminishes the background noise. Consequently, this positively impacts SNR and CNR to some extent, as demonstrated in the evaluation metrics table depicted in Table~\ref{tab2}. However,  there remains an inevitable trade-off between resolution and SNR/CNR in recurrent inference, as illustrated in Fig.~\ref{fig:fig4}. 

\begin{table}[!h]
    \centering
    \footnotesize
    \renewcommand{\arraystretch}{1.5}
    \setlength{\tabcolsep}{5pt}
    \caption{Evaluation metrics for moderate-resolution, high SNR images}
    \vspace{0pt}
    \captionsetup{skip=3pt} 
    \scalebox{1}{
    \begin{tabular}{lrr}
    \hline
    &Input&Net \\
    \hline
    EPI & 1.00 & 1.09$\pm$   0.07\\
    SNR(dB) & 77.48 $\pm$ 4.85 & 89.09 $\pm$   5.34\\
    CNR(dB) & 9.53 $\pm$ 0.55 & 9.73 $\pm$   0.68\\
    ENL & 161.25 $\pm$ 61.93 & 518.51 $\pm$ 317.83 \\
    \hline
    \end{tabular}} 
    \label{tab2}
\end{table}

Secondly, we employ $model_r$ to demonstrate the recursive generation of high-resolution images, as depicted in Fig.~\ref{fig:fig4}. This generation method empowers practitioners to choose the appropriate level of resolution enhancement to ensure the best overall resolution and SNR. The images labeled Net\_1 showcase the end-to-end outputs of the network, utilizing only the low-resolution images (`Input') as inputs. Following that, the images labeled Net\_2 were generated by taking a weighted combination of Input and Net\_1 as inputs, resulting in even higher resolution while maintaining acceptable levels of SNR and CNR. A-line profiles presented in Figs.~\ref{fig:fig4}e-h display the normalized intensity along the imaging depth, illustrating the varying levels of resolution improvement achieved by Net\_1 and Net\_2 in comparison to the `Input'. Notably, the sharper and narrower peaks observed in the A-line profiles of Net\_2 indicating its superior resolution improvement over Net\_1. However, this improvement comes at the expense of reduced SNR and CNR, signifying a trade-off for enhanced resolution. This compromise arises due to an inherent imbalance between the degree of newly introduced high-frequency noise and the noise suppression capability of the FS prior.

\begin{figure*}[ht]
\centering
\includegraphics[width=\textwidth]{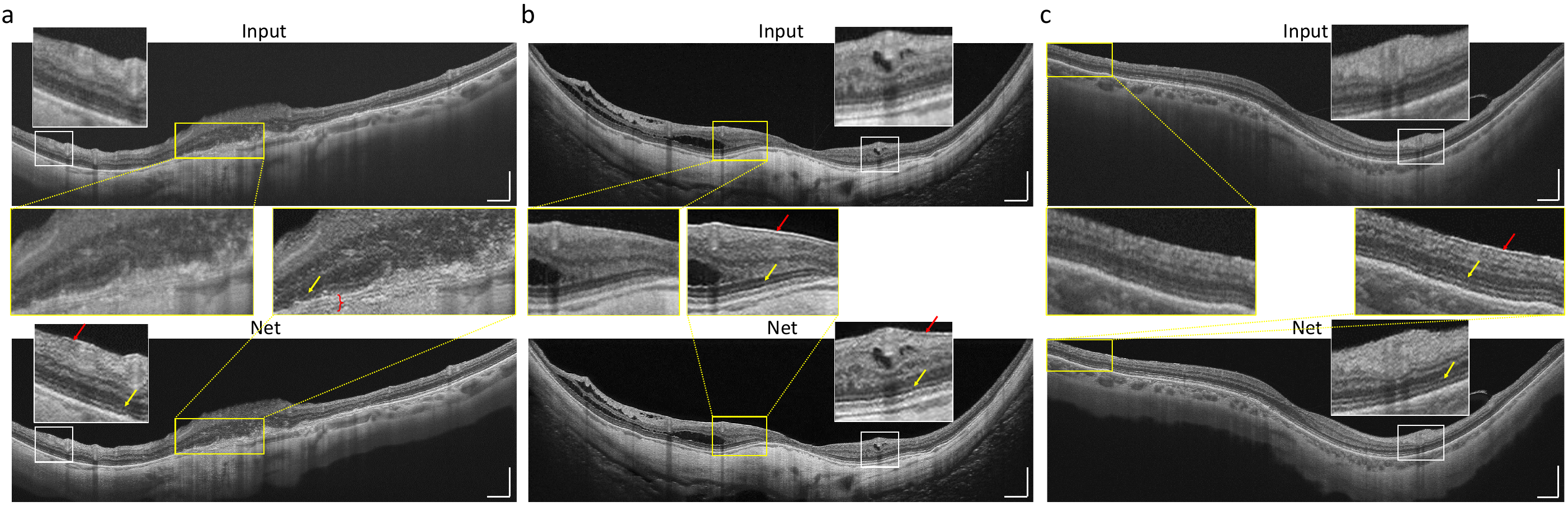}
\caption{
\begin{small}
    \textbf{Validating generalization capability of a model trained on dataset 1 (Spectralist retina images) by applying it to dataset 3 (SVision retina images) with a initially resolution of $\sim$ 3.5 $\mu$m.} The red arrows indicate the resolution-enhanced epiretinal membrane, which usually results from thickened inner limiting membrane or membrane-like changes of posterior vitreous hyaloid, while the yellow arrows indicate the resolution-enhanced external limiting membrane. The red brace in the zoomed-in image in \textbf{a} shows newly formed choriocapillary. Scale bar: 0.5 mm
\end{small} }
\label{fig:fig6}
\end{figure*}     

\subsubsection*{Validation of cross-system generalization capability on initially high-resolution retina images}
The optical axial resolutions of the validation datasets used thus far are approximately 7-8 $\mu m$. Fig.~\ref{fig:fig6} demonstrates the capacity of our method to further enhance the resolution of initially high-resolution images, revealing even finer structural details. The resolution of all input images is \(\sim \) 3.5 $\mu m$ in tissue, which is nearly the highest resolution available in commercial OCT systems used in ophthalmology. We employ $model_r$, tailored for recurrent inference, to enhance the input images. By analyzing the EPI values tabulated in Table~\ref{tab3} and examining the detailed views provided in Figs.~\ref{fig:fig6}a-c, it is easy to tell that resolution enhancement in Net images is both effective and apparent. Retina layers are more discernible, layer boundaries and lesion structures are more distinct. In terms of evaluation metrics, it is also intriguing to observe that despite the input images already having high SNR and CNR, these metrics improve further in Net images. However, the enhancement of the smoothness of the background, as represented by ENL, is somewhat limited in comparison to images achieved with moderate-resolution (Table~\ref{tab2}). In terms of inference speed, a lateral scan with a wide field of 12mm and comprising 512 (axial) × 1200 (lateral) pixels has an average processing time of 0.11 ± 0.003s when utilizing an NVidia A100 GPU, allowing for real-time reconstruction.

\begin{table}[h]
    \centering
    \footnotesize
    \renewcommand{\arraystretch}{1.5}
    \setlength{\tabcolsep}{5pt}
    \caption{Evaluation metrics for high-resolution, high SNR images}
    \vspace{0pt}
    \captionsetup{skip=3pt} 
    \scalebox{1}{
    \begin{tabular}{lrr}
    \hline
    &Input&Net \\
    \hline
    EPI & 1.00 & 1.07 $\pm$   0.05\\
    SNR(dB) & 74.34 $\pm$ 4.98 & 85.94 $\pm$   3.77\\
    CNR(dB) & 9.01 $\pm$ 0.80 & 10.24 $\pm$   0.71\\
    ENL & 157.68 $\pm$ 108.24 & 297.50 $\pm$ 171.52 \\
    \hline
    \end{tabular}} 
    \label{tab3}
\end{table}
 
\begin{figure*}[ht]
\centering
\includegraphics[width=\textwidth]{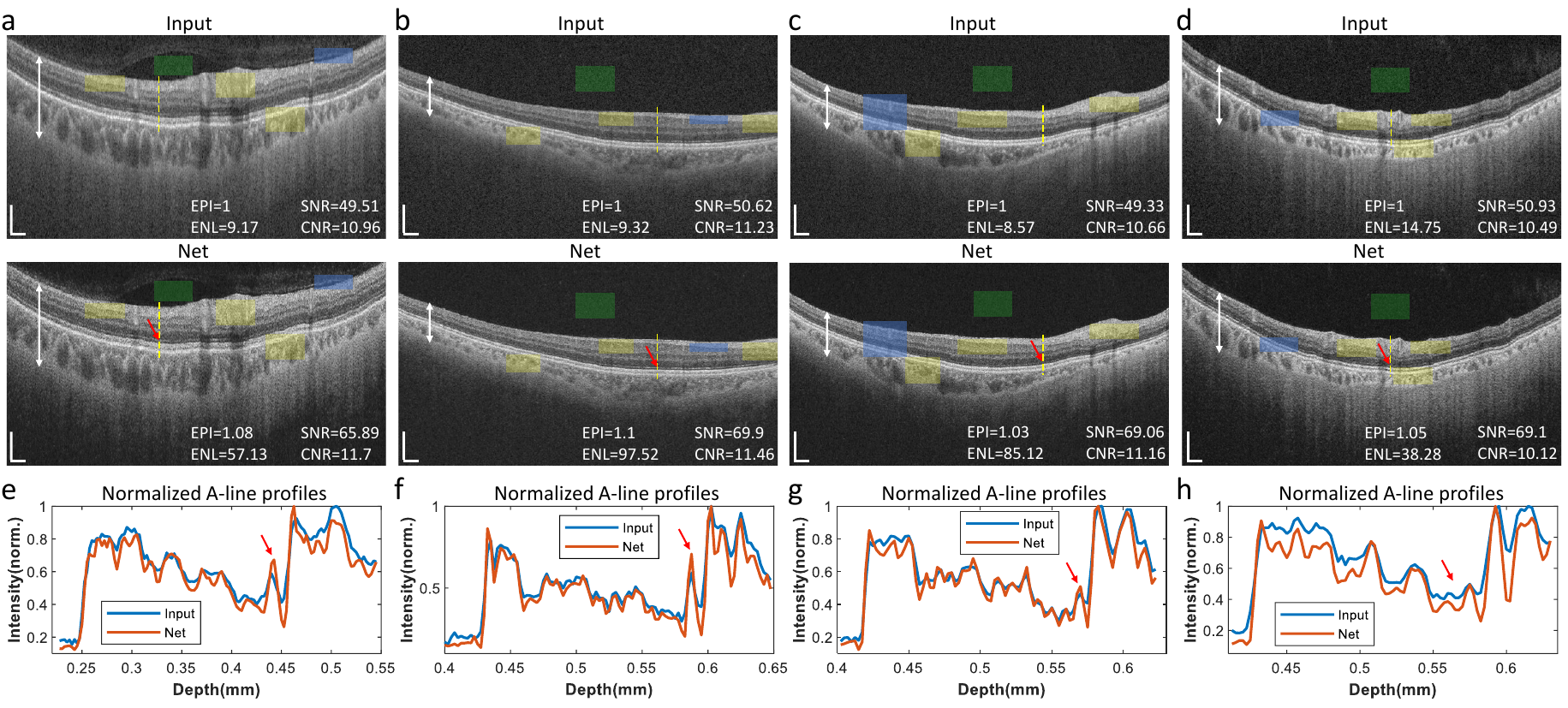}
\caption{
\begin{small}
    \textbf{Testing the robustness of $model_r$ on dataset 4 (SVision retina images) under high noise level conditions.}
 \textbf{a-d}, Input: retina OCT images extracted from dataset 4, Net: the corresponding Net=$model_r$(Input). Green patches highlight the selected background regions; blue patches indicate areas designated for EPI calculation; yellow patches mark regions chosen for CNR calculation; white double arrows define the depth ranges selected for SNR calculation.  \textbf{e-h}, normalized A-line profiles of the chosen ranges, as indicated by red dashed lines in \textbf{a-c}, respectively. Scale bar: 0.5 mm.
\end{small} }
\label{fig:fig7}
\end{figure*} 

\subsubsection*{Robustness performance}
In order to gauge the robustness of our method, we gather noisy retina images from 3D volumetric scanning dataset. These frames are obtained by averaging 2 B-frames acquired from the same location, which is notably fewer than the usual 16 or 32 averaging employed in 2D scanning. The initial axial resolution of input images is \(\sim \) 3.5 $\mu m$. By evaluating the calculated ENL values within the ROIs (green patches) shown in Fig.~\ref{fig:fig7}, we discern that the background smoothness in the `Net' images is considerably higher than that of the `Input' images. Additionally, a comparison of SNR values also revealed a substantial noise reduction in the `Net' images, averaging nearly 20 dB higher (Table~\ref{tab4}). Both the visual assessment and the improvement in evaluation metrics demonstrate that the FS prior not only suppresses noise generated during reconstruction but also that which is already present in the input images. 

\begin{table}[!h]
    \centering
    \footnotesize
    \renewcommand{\arraystretch}{1.5}
    \setlength{\tabcolsep}{5pt}
    \caption{Evaluation metrics for high-resolution, low SNR images}
    \vspace{0pt}
    \captionsetup{skip=3pt} 
    \scalebox{1}{
    \begin{tabular}{lrr}
    \hline
    &Input&Net \\
    \hline
    EPI & 1.00 & 1.08$\pm$   0.06\\
    SNR(dB) & 47.35 $\pm$ 2.02 & 66.16 $\pm$   2.05\\
    CNR(dB) & 7.90 $\pm$ 0.53 & 8.73 $\pm$   0.81\\
    ENL & 10.21 $\pm$ 2.17 & 94.57 $\pm$ 42.51 \\
    \hline
    \end{tabular}} 
    \label{tab4}
\end{table}

As shown in Fig.~\ref{fig:fig7} and summarized in Table~\ref{tab4}, despite the noise suppression, our method still achieves certain degrees of improvement in resolution, even in the presence of high noise level. This improvement has been evidenced in the EPI values, with most of the `Net' images showing an EPI greater than 1, indicating sharper layer boundaries compared to the inputs. Additionally, this improvement is also apparent in the displayed A-line profiles, where the ELM, marked by red arrows in both the intensity images and A-line profiles, appears considerably more pronounced in the `Net' images. Another interesting observation is that the speckle size in the `Net' images appears coarser compared to the `Input' images, resulting in slightly lower contrast in some `Net' images (Fig.~\ref{fig:fig7}d). This trend remains consistent across the perceptual study conducted in collaboration with optical experts. This phenomenon has also been reported in a previous study \citep{Wit2021}, which utilizes a traditional machine learning approach for axial resolution enhancement. It is likely attributed to the relatively low SNR of a single speckle, which consists of a combination of multiple unaligned sub-resolution reflectors.

\begin{figure}[!h]
\centering
\includegraphics[scale=.5]{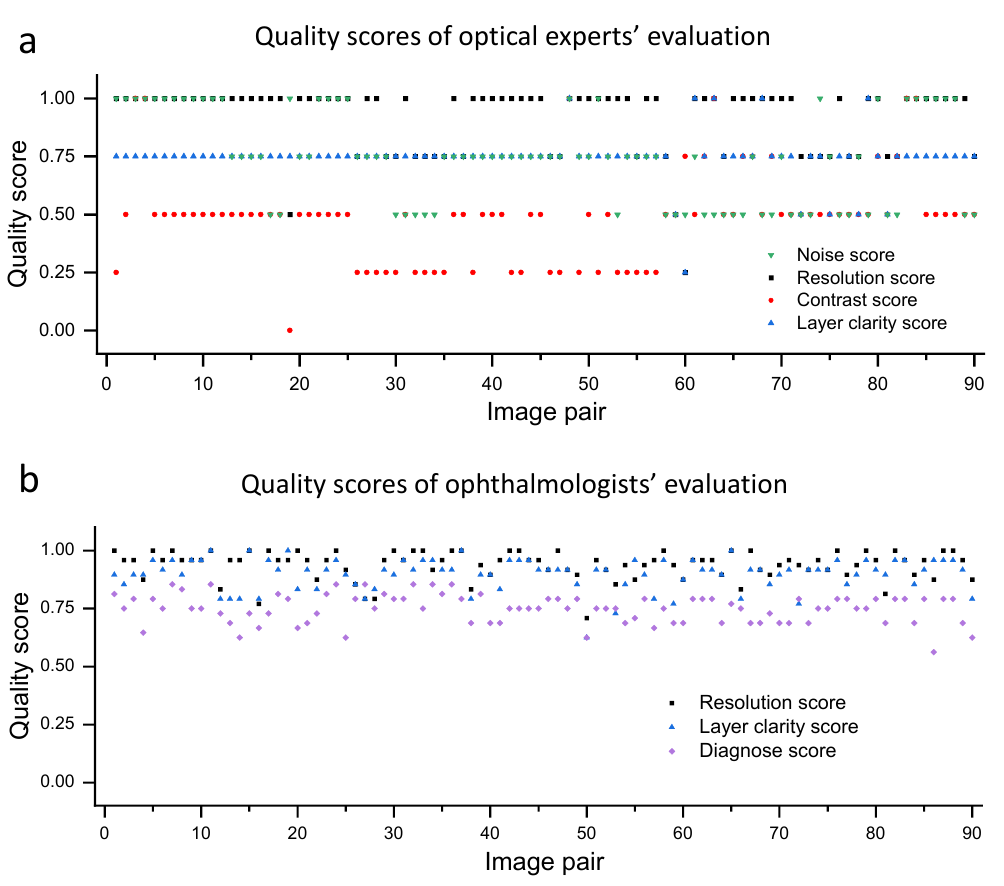}
\caption{
\begin{small}
    \textbf{Statistical analysis results of perceptual study, conducted by biophotonics imaging experts and Ophthalmologists.} \textbf{a}, Quality scores for all evaluation criteria as rated by the optical experts. \textbf{b}, Quality scores for all evaluation criteria as rated by the ophthalmologists.
\end{small} }
\label{fig:percep}
\end{figure}

\subsubsection*{Perceptual study}
Quality scores achieved by optical experts and ophthalmologists, as presented in Figs.~\ref{fig:percep}a and \ref{fig:percep}b,  demonstrate the promising performance of the proposed method in terms of visual assessment. The majority of  the calculated quality scores are equal to or greater than 0.5, indicating the superiority of the network-generated images. According to the statistical analysis summarized in Tables~\ref{tab7} and \ref{tab8}, optical experts give a highly favorable assessment of the image resolution, awarding it a score of 0.919, which suggest a significant improvement in image clarity and structural details. Both layer clarity and noise scores are above 0.5, indicating that optical experts perceives not only enhanced resolution but also reduced noise levels to some extent. The contrast score, while lower at 0.473, is still quite close to 0.5. As for ophthalmologists,  they assign very high scores to both resolution and layer clarity, and the remaining diagnostic scores, representing the ability to accurately diagnose abnormality from the images, are above 0.7. It should be emphasized that all ophthalmologists have been shown examples before the evaluation, while optical experts undergo the evaluation without any prior hints, which explains why the scores for resolution and layer clarity are not as high as those rated by the ophthalmologists.

\begin{table}[h]
    \centering
    \renewcommand{\arraystretch}{1.5}
    \setlength{\tabcolsep}{5pt}
    \caption{\label{tab7} Statistical results of optical experts' evaluation}
    \vspace{0pt}
    \captionsetup{skip=3pt} 
    \scalebox{0.8}{
    \begin{tabular}{lrrr}
    \hline
     & N total & Sum & Mean $\pm$ std \\
    \hline
    Resolution score & 90 & 82.75 & 0.919 $\pm$ 0.144 \\
    Contrast score & 90 & 43.75 & 0.486 $\pm$ 0.199 \\
    Layer clarity score & 90 & 67.00 & 0.744 $\pm$ 0.099 \\
    Noise score & 90 & 67.25& 0.747 $\pm$ 0.196 \\
    \hline
    \end{tabular}
    } 
\end{table}

\begin{table}[h]
    \centering
    \renewcommand{\arraystretch}{1.5}
    \setlength{\tabcolsep}{5pt}
    \caption{\label{tab8} Statistical results of ophthalmologists' evaluation}
    \vspace{0pt}
    \captionsetup{skip=3pt} 
    \scalebox{0.8}{
    \begin{tabular}{lrrr}
    \hline
     & N total & Sum & Mean $\pm$ std \\
    \hline
    Resolution score & 90 & 84.40 & 0.938 $\pm$ 0.061 \\
    Layer clarity score & 90 & 80.85 & 0.898 $\pm$ 0.068 \\
    Noise score & 90 & 67.42& 0.749 $\pm$ 0.063 \\
    \hline
    \end{tabular}
    } 
\end{table}

Specific score values for each evaluation criterion achieved by all evaluation groups are presented in Supplementary Fig.~\ref{fig:pecp_optics} to Supplementary Fig.~\ref{fig:pecp_senior}. Supplementary Fig.~\ref{fig:pecp_optics} displays score values given by two optical experts. It is evident  that most scores for resolution, layer clarity and noise are equal to or greater than 0.75, while contrast has a main value of 0.5. Supplementary Fig.~\ref{fig:pecp_stud} shows score values assigned by two ophthalmology students, and the corresponding statistical results. Apart from giving high scores for resolution and layer clarity,  ophthalmology students also rate the diagnosis score high, indicating that network-generated images can assist with diagnosis. Supplementary Fig.~\ref{fig:pecp_junior} displays the scores given by two ophthalmology residents. Although their scores for the three criteria are lower than those of the students, they still rate highly for resolution and layer clarity.  One resident believes that AI is very helpful for diagnosis, while the other think that AI performed similarly to the inputs.  Supplementary Fig.~\ref{fig:pecp_intermediate} displays scores given by two ophthalmologists with intermediate professional titles. The score distribution is similar to that of residents, except that the diagnosis score has a dominant value of 0.75, and the resolution scores are slightly lower.  Supplementary Fig.~\ref{fig:pecp_senior} shows the scores given by three professors of ophthalmology. In  Supplementary Fig.~\ref{fig:pecp_senior}a, the scores are distributed almost evenly between 0.5 and 1, with two outliers below 0.5. In Supplementary Fig.~\ref{fig:pecp_senior}b, although the three scores dropped significantly, the lowest score remains above 0.5, indicating that senior ophthalmologists acknowledge the improvements brought by AI.

Furthermore, in our perceptual study, we find that optical imaging experts concur that neural network-generated images exhibit a higher SNR when the input images undergo fewer instances of frame averaging. Nevertheless, when the input images are the result of 16 or 32 times averaging, the experts deem the SNR values of the neural network-generated images and the inputs to be comparable. This discovery aligns with the outcomes of the validation experiments mentioned earlier.

\subsubsection*{Validation of cross-tissue generalization capability on swine esophagus images}
\begin{figure*}[h]
    \centering
    \includegraphics[width=\textwidth]{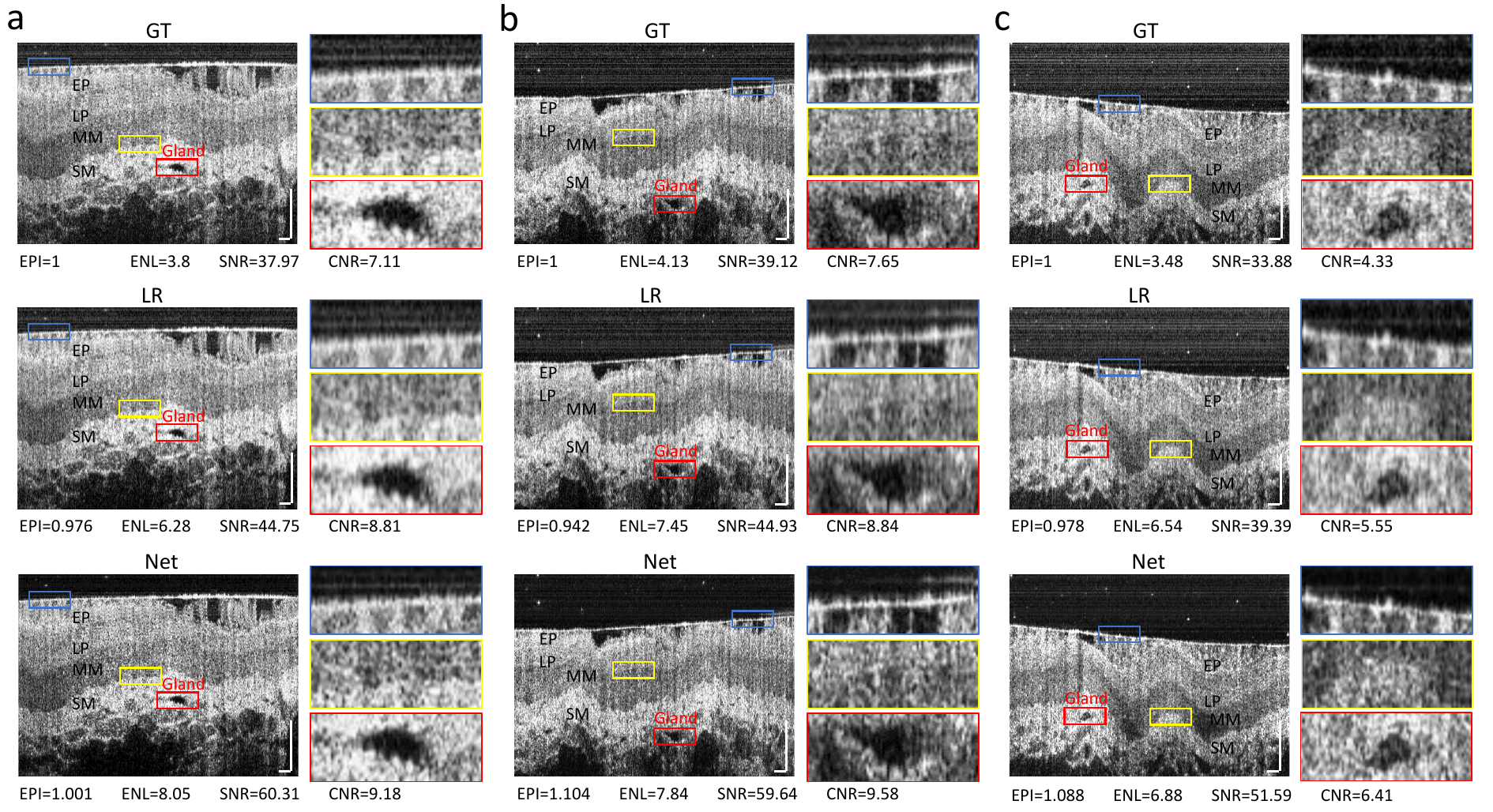}
\caption{
\begin{small}
\textbf{Generalization results of swine esophagus mucosa.} Ground Truth (GT): Ultrahigh-resolution images feature an axial resolution of 2.3 $\mu m$. Low-resolution images (LR) are obtained by convolving the GT with a PSF shown in Supplementary Fig. \ref{fig:psf}a. Reconstructed high-resolution images (Net) are generated by the network using LR as inputs. The blue boxes indicate the ROI for EPI calculation. EPI of LR and Net is calculated relative to the GT. ENL is computed from the FS region above the tissue surface, which is also used as the background for SNR and CNR calculation. The entire region below the tissue surface is selected as the ROI for SNR and CNR calculation. Yellow boxes delineate the boundaries of different layers, while red boxes indicate areas of low-signal glands within the submucosal layer. EP: Epithelium, LP: Lamina Propria, MM: Muscular Mucosa, SM: Submucosa. All scale bars: 100 $\mu m$.
\end{small} }
    \label{fig:eso}
\end{figure*}

To evaluate the model's capacity to apply knowledge to various tissues, we perform a test on swine esophagus mucosa OCT images (dataset 5 in Supplementary Table~\ref{tab5}), using $model_s$ that is trained on human retinal images. The testing data is acquired from a home-built ultrahigh resolution OCT system. To effectively demonstrate the model's capability and accuracy in cross-tissue generalization, we utilize the original ultrahigh resolution (2.6 $\mu m$) images as GT, and synthesize the low resolution inputs for the model. The brightness of the low resolution inputs is adjusted to 1/3 of the original level to align with the training data. Results from three examples are presented in Fig.~\ref{fig:eso}. The zoomed-in figures in the red and yellow boxes clearly show that the resolution and layer clarity of `Net' images are both higher than LR images. In terms of speckle structure, the speckle pattern and size of `Net' images are more similar to GT compared to LR images. Additionally, the speckle size of `Net' images is finer than that of LR images, which is closely related to the spatial resolution \citep{Wit2021}. In terms of layer clarity, the boundaries between MM and SM, LP and MM zoomed in red boxes of `Net' images are clearer than those of the LR images. Specifically, the clarity of the boundaries of the glands shown in yellow boxes within the submucosa layers in the `Net'
images has significantly improved compared to the LR images. In terms of quantifying assessment, for EPI, GT images are used as a reference and compared to LR and `Net' images. We can observe that the EPI of LR images are all smaller than 1, which aligns with our observation that tissue surfaces of LR images are blurrier than GT images. Conversely, the EPI of `Net' images are all greater than 1, which also aligns with our observation that the sharpness of the tissue surfaces in the `Net' images is very similar to the GT images. Additionally, the background of the `Net' images is much cleaner than the GT, both of which contribute to the higher EPI values. For ENL, `Net' images have the largest values resulting from the very smoothness of the free space region, i.e. the background. For both SNR and CNR, `Net' images achieve the best performance due to the effective suppression of the background noise. Along with ENL, this can be attributed to the powerful FS prior. It is also worth noting that the blurry effect experienced by LR images leads to better performance on ENL, SNR and CNR due to the smaller standard deviation of the background. 

\section{Conclusions}
The axial resolution of OCT is primarily determined and constrained by the center wavelength and spectral bandwidth of the light source employed. However, the utilization of broadband lasers to enhance resolution comes with a substantial hardware cost. Moreover, these lasers necessitate high-quality and aberration-free interferometers, coupled with expensive detectors. In a bid to further extend the capabilities of current hardware, contemporary researchers leverage computational techniques to reconstruct higher resolution images from acquired measurements. Nevertheless, there exist noteworthy limitations in the current methodologies: 1) sluggish reconstruction speed; 2) reliance on high-resolution images or raw spectral data; 3) restricted generalization ability. 

In this paper, we introduce a novel and robust self-supervised learning approach to enhance the axial resolution of OCT, specifically tailored for situations where only OCT intensity images are available. This approach not only allows for a higher resolution with existing hardware, but also has the potential to use compact and affordable narrow band light sources as a replacement for bulky and expensive broadband ones while maintaining the same level of resolution. The key contributions of our work to the OCT community can be summarized as follows:
1) \textbf{Robustness and Effectiveness:} When input images are noisy, our method exhibits denoising capabilities armed by the FS prior, and maintaining its efficacy across various imaging scenarios. Contrasted with conventional machine learning techniques, our approach does not directly estimate 
sample reflectivity $r_S$ A-line by A-line, but learns an inverse mapping and generates $r_S$ images frame-by-frame, resulting in real-time reconstruct speed.
2) \textbf{Self-Supervised Learning:} Unlike conventional LR-HR pair-based supervised methods, our approach leverages the physical model of OCT and incorporates the guidance of EI and FS priors for learning. This enables us to achieve performance that is comparable to fully-supervised learning methods. Through extensive experimentation on simulated data, diverse human retina samples, and the swine esophagus images, we showcase the efficacy and superiority of our proposed method. In comparison to the traditional RL method, our approach yields images with more authentic speckle patterns and intensities. 
3) \textbf{Wide Applicability:} Our work is zero-shot and trained exclusively on LR images without supervision from HR counterparts. Therefore, models trained on retina data can smoothly extend to various tissue types acquired from different systems, requiring only simple intensity and digital resolution adjustments. This makes our method easily adaptable to other imaging modalities such as MR, CT, ultrasound, and many other optical imaging techniques, with the potential to influence and advance the resolution of multiple domains in medical imaging.


\section*{CRediT authorship contribution statement}
\noindent \textbf{Kaiyan Li:} Conceptualization, Methodology, Data preparation, Formal analysis, Software, Validation, Visualization, Writing - original draft and Funding acquisition;
\textbf{Jingyuan Yang:} Data preparation, Validation and Writing - review \& editing;
\textbf{Wenxuan Liang:} Methodology and Validation;
\textbf{Xingde Li:} Data preparation and Validation;
\textbf{Chenxi Zhang:} Validation;
\textbf{Lulu Chen:} Validation;
\textbf{Chan Wu:} Validation;
\textbf{Xiao Zhang:} Validation;
\textbf{Zhiyan Xu:} Validation;
\textbf{Yuelin Wang:} Validation;
\textbf{Lihui Meng:} Validation;
\textbf{Yue Zhang:}  Writing - review \& editing;
\textbf{Youxin Chen:} Validation, Funding acquisition, Supervision and Project administration;
\textbf{S. Kevin Zhou:} Conceptualization, Methodology, Funding acquisition, Project administration, Supervision and Writing - review \& editing.

\section*{Declaration of competing interest}
The authors declare that they have no known competing financial interests or personal relationships that could have appeared to
influence the work reported in this paper. 

\section*{Data availability}
The main data supporting the results in this study are available within the paper and its Supplementary Information. Due to the consideration of patients privacy, the raw and analysed datasets generated during the study are not appropriate to be publicly shared right now, yet they are available for research purposes from the corresponding authors on reasonable request. 

\section*{Code availability}
The code base for the deep-learning framework were adapted from https://github.com/edongdongchen/EI. Customized changes were made accordingly in aspects of OCT imaging, priors, loss functions and data preparations. All experiments and implementation details are described in sufficient detail in the Methods and in the Supplementary Figures to enable independent replication with non-proprietary libraries. All custom MATLAB and python scripts used to pre-process and train data are available from the authors upon request.

\section*{Acknowledgments}
We thank Jun Li(USTC) and Zikang Xu(USTC) for their kind technical support. We also appreciate the helpful discussion provided by Qiuli Wang (USTC) and Mengyuan Wang (SJTU).This research was supported by the National High Level Hospital Clinical Research Funding 2022-PUMCH-B-101, the Natural Science Foundation of China 62271465, Open Fund Project of Guangdong Academy of Medical Sciences of China YKY-KF202206, Jiangsu Funding Program for Excellent Postdoctoral Talent and China Postdoctoral Science Foundation funded project 2023M733428.

\bibliographystyle{model2-names.bst}\biboptions{authoryear}

\begin{thebibliography}{54}
\expandafter\ifx\csname natexlab\endcsname\relax\def\natexlab#1{#1}\fi
\providecommand{\url}[1]{\texttt{#1}}
\providecommand{\href}[2]{#2}
\providecommand{\path}[1]{#1}
\providecommand{\DOIprefix}{doi:}
\providecommand{\ArXivprefix}{arXiv:}
\providecommand{\URLprefix}{URL: }
\providecommand{\Pubmedprefix}{pmid:}
\providecommand{\doi}[1]{\href{http://dx.doi.org/#1}{\path{#1}}}
\providecommand{\Pubmed}[1]{\href{pmid:#1}{\path{#1}}}
\providecommand{\bibinfo}[2]{#2}
\ifx\xfnm\relax \def\xfnm[#1]{\unskip,\space#1}\fi
\bibitem[{Agrawal et~al.(2017)Agrawal, Pfefer, Woolliams, Tomlins and Nehmetallah}]{agrawal2017methods}
\bibinfo{author}{Agrawal, A.}, \bibinfo{author}{Pfefer, T.J.}, \bibinfo{author}{Woolliams, P.D.}, \bibinfo{author}{Tomlins, P.H.}, \bibinfo{author}{Nehmetallah, G.}, \bibinfo{year}{2017}.
\newblock \bibinfo{title}{Methods to assess sensitivity of optical coherence tomography systems}.
\newblock \bibinfo{journal}{Biomedical Optics Express} \bibinfo{volume}{8}, \bibinfo{pages}{902--917}.
\bibitem[{Azizi et~al.(2021)Azizi, Mustafa, Ryan, Beaver, Freyberg, Deaton, Loh, Karthikesalingam, Kornblith, Chen et~al.}]{azizi2021big}
\bibinfo{author}{Azizi, S.}, \bibinfo{author}{Mustafa, B.}, \bibinfo{author}{Ryan, F.}, \bibinfo{author}{Beaver, Z.}, \bibinfo{author}{Freyberg, J.}, \bibinfo{author}{Deaton, J.}, \bibinfo{author}{Loh, A.}, \bibinfo{author}{Karthikesalingam, A.}, \bibinfo{author}{Kornblith, S.}, \bibinfo{author}{Chen, T.}, et~al., \bibinfo{year}{2021}.
\newblock \bibinfo{title}{Big self-supervised models advance medical image classification}, in: \bibinfo{booktitle}{Proceedings of the IEEE/CVF International Conference on Computer Vision}, pp. \bibinfo{pages}{3478--3488}.
\bibitem[{Bai et~al.(2019)Bai, Chen, Tarroni, Duan, Guitton, Petersen, Guo, Matthews and Rueckert}]{bai2019self}
\bibinfo{author}{Bai, W.}, \bibinfo{author}{Chen, C.}, \bibinfo{author}{Tarroni, G.}, \bibinfo{author}{Duan, J.}, \bibinfo{author}{Guitton, F.}, \bibinfo{author}{Petersen, S.E.}, \bibinfo{author}{Guo, Y.}, \bibinfo{author}{Matthews, P.M.}, \bibinfo{author}{Rueckert, D.}, \bibinfo{year}{2019}.
\newblock \bibinfo{title}{Self-supervised learning for cardiac mr image segmentation by anatomical position prediction}, in: \bibinfo{booktitle}{Medical Image Computing and Computer Assisted Intervention--MICCAI 2019: 22nd International Conference, Shenzhen, China, October 13--17, 2019, Proceedings, Part II 22}, \bibinfo{organization}{Springer}. pp. \bibinfo{pages}{541--549}.
\bibitem[{Baumann et~al.(2019)Baumann, Merkle, Leitgeb, Augustin, Wartak, Pircher and Hitzenberger}]{baumann2019signal}
\bibinfo{author}{Baumann, B.}, \bibinfo{author}{Merkle, C.W.}, \bibinfo{author}{Leitgeb, R.A.}, \bibinfo{author}{Augustin, M.}, \bibinfo{author}{Wartak, A.}, \bibinfo{author}{Pircher, M.}, \bibinfo{author}{Hitzenberger, C.K.}, \bibinfo{year}{2019}.
\newblock \bibinfo{title}{Signal averaging improves signal-to-noise in \uppercase{OCT} images: But which approach works best, and when?}
\newblock \bibinfo{journal}{Biomedical Optics Express} \bibinfo{volume}{10}, \bibinfo{pages}{5755--5775}.
\bibitem[{Cao et~al.(2020)Cao, Yao, Koirala, Brott, Litovsky, Ling and Gan}]{Cao2020}
\bibinfo{author}{Cao, S.}, \bibinfo{author}{Yao, X.}, \bibinfo{author}{Koirala, N.}, \bibinfo{author}{Brott, B.}, \bibinfo{author}{Litovsky, S.}, \bibinfo{author}{Ling, Y.}, \bibinfo{author}{Gan, Y.}, \bibinfo{year}{2020}.
\newblock \bibinfo{title}{Super-resolution technology to simultaneously improve optical \& digital resolution of optical coherence tomography via deep learning}, in: \bibinfo{booktitle}{2020 42nd Annual International Conference of the IEEE Engineering in Medicine \& Biology Society (EMBC)}, \bibinfo{organization}{IEEE}. pp. \bibinfo{pages}{1879--1882}.
\bibitem[{Chen et~al.(2021)Chen, Tachella and Davies}]{Chen2021}
\bibinfo{author}{Chen, D.}, \bibinfo{author}{Tachella, J.}, \bibinfo{author}{Davies, M.E.}, \bibinfo{year}{2021}.
\newblock \bibinfo{title}{Equivariant imaging: Learning beyond the range space}, in: \bibinfo{booktitle}{Proceedings of the IEEE/CVF International Conference on Computer Vision}, pp. \bibinfo{pages}{4379--4388}.
\bibitem[{Chen et~al.(2019)Chen, Bentley, Mori, Misawa, Fujiwara and Rueckert}]{chen2019self}
\bibinfo{author}{Chen, L.}, \bibinfo{author}{Bentley, P.}, \bibinfo{author}{Mori, K.}, \bibinfo{author}{Misawa, K.}, \bibinfo{author}{Fujiwara, M.}, \bibinfo{author}{Rueckert, D.}, \bibinfo{year}{2019}.
\newblock \bibinfo{title}{Self-supervised learning for medical image analysis using image context restoration}.
\newblock \bibinfo{journal}{Medical Image Analysis} \bibinfo{volume}{58}, \bibinfo{pages}{101539}.
\bibitem[{De~Boer et~al.(2003)De~Boer, Cense, Park, Pierce, Tearney and Bouma}]{de2003improved}
\bibinfo{author}{De~Boer, J.F.}, \bibinfo{author}{Cense, B.}, \bibinfo{author}{Park, B.H.}, \bibinfo{author}{Pierce, M.C.}, \bibinfo{author}{Tearney, G.J.}, \bibinfo{author}{Bouma, B.E.}, \bibinfo{year}{2003}.
\newblock \bibinfo{title}{Improved signal-to-noise ratio in spectral-domain compared with time-domain optical coherence tomography}.
\newblock \bibinfo{journal}{Optics Letters} \bibinfo{volume}{28}, \bibinfo{pages}{2067--2069}.
\bibitem[{De~Fauw et~al.(2018)De~Fauw, Ledsam, Romera-Paredes, Nikolov, Tomasev, Blackwell, Askham, Glorot, O’Donoghue, Visentin et~al.}]{de2018clinically}
\bibinfo{author}{De~Fauw, J.}, \bibinfo{author}{Ledsam, J.R.}, \bibinfo{author}{Romera-Paredes, B.}, \bibinfo{author}{Nikolov, S.}, \bibinfo{author}{Tomasev, N.}, \bibinfo{author}{Blackwell, S.}, \bibinfo{author}{Askham, H.}, \bibinfo{author}{Glorot, X.}, \bibinfo{author}{O’Donoghue, B.}, \bibinfo{author}{Visentin, D.}, et~al., \bibinfo{year}{2018}.
\newblock \bibinfo{title}{Clinically applicable deep learning for diagnosis and referral in retinal disease}.
\newblock \bibinfo{journal}{Nature Medicine} \bibinfo{volume}{24}, \bibinfo{pages}{1342--1350}.
\bibitem[{De~Wit et~al.(2021)De~Wit, Angelopoulos, Kalkman and Glentis}]{Wit2021}
\bibinfo{author}{De~Wit, J.}, \bibinfo{author}{Angelopoulos, K.}, \bibinfo{author}{Kalkman, J.}, \bibinfo{author}{Glentis, G.O.}, \bibinfo{year}{2021}.
\newblock \bibinfo{title}{Fast and accurate spectral-estimation axial super-resolution optical coherence tomography}.
\newblock \bibinfo{journal}{Optics Express} \bibinfo{volume}{29}, \bibinfo{pages}{39946--39966}.
\bibitem[{Drexler et~al.(1999)Drexler, Morgner, K{\"a}rtner, Pitris, Boppart, Li, Ippen and Fujimoto}]{drexler1999vivo}
\bibinfo{author}{Drexler, W.}, \bibinfo{author}{Morgner, U.}, \bibinfo{author}{K{\"a}rtner, F.}, \bibinfo{author}{Pitris, C.}, \bibinfo{author}{Boppart, S.}, \bibinfo{author}{Li, X.}, \bibinfo{author}{Ippen, E.}, \bibinfo{author}{Fujimoto, J.}, \bibinfo{year}{1999}.
\newblock \bibinfo{title}{In vivo ultrahigh-resolution optical coherence tomography}.
\newblock \bibinfo{journal}{Optics Letters} \bibinfo{volume}{24}, \bibinfo{pages}{1221--1223}.
\bibitem[{Farsiu et~al.(2014)Farsiu, Chiu, O'Connell, Folgar, Yuan, Izatt, Toth, Group et~al.}]{Farsiu2014}
\bibinfo{author}{Farsiu, S.}, \bibinfo{author}{Chiu, S.J.}, \bibinfo{author}{O'Connell, R.V.}, \bibinfo{author}{Folgar, F.A.}, \bibinfo{author}{Yuan, E.}, \bibinfo{author}{Izatt, J.A.}, \bibinfo{author}{Toth, C.A.}, \bibinfo{author}{Group, A.R.E.D.S..A.S.D.O.C.T.S.}, et~al., \bibinfo{year}{2014}.
\newblock \bibinfo{title}{Quantitative classification of eyes with and without intermediate age-related macular degeneration using optical coherence tomography}.
\newblock \bibinfo{journal}{Ophthalmology} \bibinfo{volume}{121}, \bibinfo{pages}{162--172}.
\bibitem[{Fercher(1996)}]{fercher1996optical}
\bibinfo{author}{Fercher, A.F.}, \bibinfo{year}{1996}.
\newblock \bibinfo{title}{Optical coherence tomography}.
\newblock \bibinfo{journal}{Journal of Biomedical Optics} \bibinfo{volume}{1}, \bibinfo{pages}{157--173}.
\bibitem[{Fercher et~al.(2001)Fercher, Hitzenberger, Sticker, Zawadzki, Karamata and Lasser}]{fercher2001numerical}
\bibinfo{author}{Fercher, A.F.}, \bibinfo{author}{Hitzenberger, C.K.}, \bibinfo{author}{Sticker, M.}, \bibinfo{author}{Zawadzki, R.}, \bibinfo{author}{Karamata, B.}, \bibinfo{author}{Lasser, T.}, \bibinfo{year}{2001}.
\newblock \bibinfo{title}{Numerical dispersion compensation for partial coherence interferometry and optical coherence tomography}.
\newblock \bibinfo{journal}{Optics Express} \bibinfo{volume}{9}, \bibinfo{pages}{610--615}.
\bibitem[{Halupka et~al.(2018)Halupka, Antony, Lee, Lucy, Rai, Ishikawa, Wollstein, Schuman and Garnavi}]{halupka2018retinal}
\bibinfo{author}{Halupka, K.J.}, \bibinfo{author}{Antony, B.J.}, \bibinfo{author}{Lee, M.H.}, \bibinfo{author}{Lucy, K.A.}, \bibinfo{author}{Rai, R.S.}, \bibinfo{author}{Ishikawa, H.}, \bibinfo{author}{Wollstein, G.}, \bibinfo{author}{Schuman, J.S.}, \bibinfo{author}{Garnavi, R.}, \bibinfo{year}{2018}.
\newblock \bibinfo{title}{Retinal optical coherence tomography image enhancement via deep learning}.
\newblock \bibinfo{journal}{Biomedical 0ptics Express} \bibinfo{volume}{9}, \bibinfo{pages}{6205--6221}.
\bibitem[{Hariri et~al.(2009)Hariri, Moayed, Dracopoulos, Hyun, Boyd and Bizheva}]{hariri2009limiting}
\bibinfo{author}{Hariri, S.}, \bibinfo{author}{Moayed, A.A.}, \bibinfo{author}{Dracopoulos, A.}, \bibinfo{author}{Hyun, C.}, \bibinfo{author}{Boyd, S.}, \bibinfo{author}{Bizheva, K.}, \bibinfo{year}{2009}.
\newblock \bibinfo{title}{Limiting factors to the \uppercase{OCT} axial resolution for in-vivo imaging of human and rodent retina in the 1060nm wavelength range}.
\newblock \bibinfo{journal}{Optics Express} \bibinfo{volume}{17}, \bibinfo{pages}{24304--24316}.
\bibitem[{Hu and Rollins(2007)}]{hu2007fourier}
\bibinfo{author}{Hu, Z.}, \bibinfo{author}{Rollins, A.M.}, \bibinfo{year}{2007}.
\newblock \bibinfo{title}{Fourier domain optical coherence tomography with a linear-in-wavenumber spectrometer}.
\newblock \bibinfo{journal}{Optics Letters} \bibinfo{volume}{32}, \bibinfo{pages}{3525--3527}.
\bibitem[{Huang et~al.(1991)Huang, Swanson, Lin, Schuman, Stinson, Chang, Hee, Flotte, Gregory, Puliafito et~al.}]{huang1991optical}
\bibinfo{author}{Huang, D.}, \bibinfo{author}{Swanson, E.A.}, \bibinfo{author}{Lin, C.P.}, \bibinfo{author}{Schuman, J.S.}, \bibinfo{author}{Stinson, W.G.}, \bibinfo{author}{Chang, W.}, \bibinfo{author}{Hee, M.R.}, \bibinfo{author}{Flotte, T.}, \bibinfo{author}{Gregory, K.}, \bibinfo{author}{Puliafito, C.A.}, et~al., \bibinfo{year}{1991}.
\newblock \bibinfo{title}{Optical coherence tomography}.
\newblock \bibinfo{journal}{science} \bibinfo{volume}{254}, \bibinfo{pages}{1178--1181}.
\bibitem[{Huang et~al.(2023)Huang, Pareek, Jensen, Lungren, Yeung and Chaudhari}]{huang2023self}
\bibinfo{author}{Huang, S.C.}, \bibinfo{author}{Pareek, A.}, \bibinfo{author}{Jensen, M.}, \bibinfo{author}{Lungren, M.P.}, \bibinfo{author}{Yeung, S.}, \bibinfo{author}{Chaudhari, A.S.}, \bibinfo{year}{2023}.
\newblock \bibinfo{title}{Self-supervised learning for medical image classification: a systematic review and implementation guidelines}.
\newblock \bibinfo{journal}{NPJ Digital Medicine} \bibinfo{volume}{6}, \bibinfo{pages}{74}.
\bibitem[{Huang et~al.(2019)Huang, Lu, Shao, Ran, Zhou, Fang and Zhang}]{HuangOE2019}
\bibinfo{author}{Huang, Y.}, \bibinfo{author}{Lu, Z.}, \bibinfo{author}{Shao, Z.}, \bibinfo{author}{Ran, M.}, \bibinfo{author}{Zhou, J.}, \bibinfo{author}{Fang, L.}, \bibinfo{author}{Zhang, Y.}, \bibinfo{year}{2019}.
\newblock \bibinfo{title}{Simultaneous denoising and super-resolution of optical coherence tomography images based on generative adversarial network}.
\newblock \bibinfo{journal}{Optics Express} \bibinfo{volume}{27}, \bibinfo{pages}{12289--12307}.
\bibitem[{Huang et~al.(2020)Huang, Xia, Lu, Liu, Chen, Zhou, Fang and Zhang}]{huang2020noise}
\bibinfo{author}{Huang, Y.}, \bibinfo{author}{Xia, W.}, \bibinfo{author}{Lu, Z.}, \bibinfo{author}{Liu, Y.}, \bibinfo{author}{Chen, H.}, \bibinfo{author}{Zhou, J.}, \bibinfo{author}{Fang, L.}, \bibinfo{author}{Zhang, Y.}, \bibinfo{year}{2020}.
\newblock \bibinfo{title}{Noise-powered disentangled representation for unsupervised speckle reduction of optical coherence tomography images}.
\newblock \bibinfo{journal}{IEEE Transactions on Medical Imaging} \bibinfo{volume}{40}, \bibinfo{pages}{2600--2614}.
\bibitem[{Institute and America(2007)}]{institute2007american}
\bibinfo{author}{Institute, A.}, \bibinfo{author}{America, L.o.}, \bibinfo{year}{2007}.
\newblock \bibinfo{title}{American national standard for safe use of lasers}.
\bibitem[{Izatt et~al.(1996)Izatt, Kulkarni, Wang, Kobayashi and Sivak}]{izatt1996optical}
\bibinfo{author}{Izatt, J.A.}, \bibinfo{author}{Kulkarni, M.D.}, \bibinfo{author}{Wang, H.W.}, \bibinfo{author}{Kobayashi, K.}, \bibinfo{author}{Sivak, M.V.}, \bibinfo{year}{1996}.
\newblock \bibinfo{title}{Optical coherence tomography and microscopy in gastrointestinal tissues}.
\newblock \bibinfo{journal}{IEEE Journal of Selected topics in quantum electronics} \bibinfo{volume}{2}, \bibinfo{pages}{1017--1028}.
\bibitem[{Kermany et~al.(2018)Kermany, Goldbaum, Cai, Valentim, Liang, Baxter, McKeown, Yang, Wu, Yan et~al.}]{Kermany2018}
\bibinfo{author}{Kermany, D.S.}, \bibinfo{author}{Goldbaum, M.}, \bibinfo{author}{Cai, W.}, \bibinfo{author}{Valentim, C.C.}, \bibinfo{author}{Liang, H.}, \bibinfo{author}{Baxter, S.L.}, \bibinfo{author}{McKeown, A.}, \bibinfo{author}{Yang, G.}, \bibinfo{author}{Wu, X.}, \bibinfo{author}{Yan, F.}, et~al., \bibinfo{year}{2018}.
\newblock \bibinfo{title}{Identifying medical diagnoses and treatable diseases by image-based deep learning}.
\newblock \bibinfo{journal}{cell} \bibinfo{volume}{172}, \bibinfo{pages}{1122--1131}.
\bibitem[{Kirillov et~al.(2023)Kirillov, Mintun, Ravi, Mao, Rolland, Gustafson, Xiao, Whitehead, Berg, Lo et~al.}]{kirillov2023segment}
\bibinfo{author}{Kirillov, A.}, \bibinfo{author}{Mintun, E.}, \bibinfo{author}{Ravi, N.}, \bibinfo{author}{Mao, H.}, \bibinfo{author}{Rolland, C.}, \bibinfo{author}{Gustafson, L.}, \bibinfo{author}{Xiao, T.}, \bibinfo{author}{Whitehead, S.}, \bibinfo{author}{Berg, A.C.}, \bibinfo{author}{Lo, W.Y.}, et~al., \bibinfo{year}{2023}.
\newblock \bibinfo{title}{Segment anything}.
\newblock \bibinfo{journal}{arXiv preprint arXiv:2304.02643} .
\bibitem[{Klein and Huber(2017)}]{klein2017high}
\bibinfo{author}{Klein, T.}, \bibinfo{author}{Huber, R.}, \bibinfo{year}{2017}.
\newblock \bibinfo{title}{High-speed \uppercase{OCT} light sources and systems}.
\newblock \bibinfo{journal}{Biomedical Optics Express} \bibinfo{volume}{8}, \bibinfo{pages}{828--859}.
\bibitem[{Lazaridis et~al.(2021)Lazaridis, Lorenzi, Mohamed-Noriega, Aguilar-Munoa, Suzuki, Nomoto, Ourselin, Garway-Heath, Crabb, Bunce et~al.}]{lazaridis2021oct}
\bibinfo{author}{Lazaridis, G.}, \bibinfo{author}{Lorenzi, M.}, \bibinfo{author}{Mohamed-Noriega, J.}, \bibinfo{author}{Aguilar-Munoa, S.}, \bibinfo{author}{Suzuki, K.}, \bibinfo{author}{Nomoto, H.}, \bibinfo{author}{Ourselin, S.}, \bibinfo{author}{Garway-Heath, D.F.}, \bibinfo{author}{Crabb, D.P.}, \bibinfo{author}{Bunce, C.}, et~al., \bibinfo{year}{2021}.
\newblock \bibinfo{title}{\uppercase{OCT} signal enhancement with deep learning}.
\newblock \bibinfo{journal}{Ophthalmology Glaucoma} \bibinfo{volume}{4}, \bibinfo{pages}{295--304}.
\bibitem[{Lee et~al.(2017)Lee, Tyring, Deruyter, Wu, Rokem and Lee}]{lee2017deep}
\bibinfo{author}{Lee, C.S.}, \bibinfo{author}{Tyring, A.J.}, \bibinfo{author}{Deruyter, N.P.}, \bibinfo{author}{Wu, Y.}, \bibinfo{author}{Rokem, A.}, \bibinfo{author}{Lee, A.Y.}, \bibinfo{year}{2017}.
\newblock \bibinfo{title}{Deep-learning based, automated segmentation of macular edema in optical coherence tomography}.
\newblock \bibinfo{journal}{Biomedical Optics Express} \bibinfo{volume}{8}, \bibinfo{pages}{3440--3448}.
\bibitem[{Lee et~al.(2023)Lee, Nam, Seok, Oh, Kim and Yoo}]{Lee2023}
\bibinfo{author}{Lee, W.}, \bibinfo{author}{Nam, H.S.}, \bibinfo{author}{Seok, J.Y.}, \bibinfo{author}{Oh, W.Y.}, \bibinfo{author}{Kim, J.W.}, \bibinfo{author}{Yoo, H.}, \bibinfo{year}{2023}.
\newblock \bibinfo{title}{Deep learning-based image enhancement in optical coherence tomography by exploiting interference fringe}.
\newblock \bibinfo{journal}{Communications Biology} \bibinfo{volume}{6}, \bibinfo{pages}{464}.
\bibitem[{Li et~al.(2019)Li, Liang, Mavadia-Shukla, Park, Li, Yuan, Wan and Li}]{Li2019}
\bibinfo{author}{Li, K.}, \bibinfo{author}{Liang, W.}, \bibinfo{author}{Mavadia-Shukla, J.}, \bibinfo{author}{Park, H.C.}, \bibinfo{author}{Li, D.}, \bibinfo{author}{Yuan, W.}, \bibinfo{author}{Wan, S.}, \bibinfo{author}{Li, X.}, \bibinfo{year}{2019}.
\newblock \bibinfo{title}{Super-achromatic optical coherence tomography capsule for ultrahigh-resolution imaging of esophagus}.
\newblock \bibinfo{journal}{Journal of Biophotonics} \bibinfo{volume}{12}, \bibinfo{pages}{e201800205}.
\bibitem[{Liang et~al.(2020)Liang, Liu, Chen, Xie, Lee, Liu and Lee}]{Liang2020}
\bibinfo{author}{Liang, K.}, \bibinfo{author}{Liu, X.}, \bibinfo{author}{Chen, S.}, \bibinfo{author}{Xie, J.}, \bibinfo{author}{Lee, W.Q.}, \bibinfo{author}{Liu, L.}, \bibinfo{author}{Lee, H.K.}, \bibinfo{year}{2020}.
\newblock \bibinfo{title}{Resolution enhancement and realistic speckle recovery with generative adversarial modeling of micro-optical coherence tomography}.
\newblock \bibinfo{journal}{Biomedical Optics Express} \bibinfo{volume}{11}, \bibinfo{pages}{7236--7252}.
\bibitem[{Ling et~al.(2020)Ling, Wang, Gan, Yao, Schmetterer, Zhou and Su}]{ling2020beyond}
\bibinfo{author}{Ling, Y.}, \bibinfo{author}{Wang, M.}, \bibinfo{author}{Gan, Y.}, \bibinfo{author}{Yao, X.}, \bibinfo{author}{Schmetterer, L.}, \bibinfo{author}{Zhou, C.}, \bibinfo{author}{Su, Y.}, \bibinfo{year}{2020}.
\newblock \bibinfo{title}{Beyond fourier transform: super-resolving optical coherence tomography}.
\newblock \bibinfo{journal}{arXiv preprint arXiv:2001.03129} .
\bibitem[{Liu et~al.(2011)Liu, Gardecki, Nadkarni, Toussaint, Yagi, Bouma and Tearney}]{liu2011imaging}
\bibinfo{author}{Liu, L.}, \bibinfo{author}{Gardecki, J.A.}, \bibinfo{author}{Nadkarni, S.K.}, \bibinfo{author}{Toussaint, J.D.}, \bibinfo{author}{Yagi, Y.}, \bibinfo{author}{Bouma, B.E.}, \bibinfo{author}{Tearney, G.J.}, \bibinfo{year}{2011}.
\newblock \bibinfo{title}{Imaging the subcellular structure of human coronary atherosclerosis using micro--optical coherence tomography}.
\newblock \bibinfo{journal}{Nature Medicine} \bibinfo{volume}{17}, \bibinfo{pages}{1010--1014}.
\bibitem[{Liu et~al.(2015)Liu, Chen, Cui, Yu and Liu}]{Liu2015}
\bibinfo{author}{Liu, X.}, \bibinfo{author}{Chen, S.}, \bibinfo{author}{Cui, D.}, \bibinfo{author}{Yu, X.}, \bibinfo{author}{Liu, L.}, \bibinfo{year}{2015}.
\newblock \bibinfo{title}{Spectral estimation optical coherence tomography for axial super-resolution}.
\newblock \bibinfo{journal}{Optics Express} \bibinfo{volume}{23}, \bibinfo{pages}{26521--26532}.
\bibitem[{Ma et~al.(2018)Ma, Chen, Zhu, Cheng, Xiang and Shi}]{Ma2018}
\bibinfo{author}{Ma, Y.}, \bibinfo{author}{Chen, X.}, \bibinfo{author}{Zhu, W.}, \bibinfo{author}{Cheng, X.}, \bibinfo{author}{Xiang, D.}, \bibinfo{author}{Shi, F.}, \bibinfo{year}{2018}.
\newblock \bibinfo{title}{Speckle noise reduction in optical coherence tomography images based on edge-sensitive cgan}.
\newblock \bibinfo{journal}{Biomedical Optics Express} \bibinfo{volume}{9}, \bibinfo{pages}{5129--5146}.
\bibitem[{Nassif et~al.(2004)Nassif, Cense, Park, Yun, Chen, Bouma, Tearney and de~Boer}]{nassif2004vivo}
\bibinfo{author}{Nassif, N.}, \bibinfo{author}{Cense, B.}, \bibinfo{author}{Park, B.H.}, \bibinfo{author}{Yun, S.H.}, \bibinfo{author}{Chen, T.C.}, \bibinfo{author}{Bouma, B.E.}, \bibinfo{author}{Tearney, G.J.}, \bibinfo{author}{de~Boer, J.F.}, \bibinfo{year}{2004}.
\newblock \bibinfo{title}{In vivo human retinal imaging by ultrahigh-speed spectral domain optical coherence tomography}.
\newblock \bibinfo{journal}{Optics Letters} \bibinfo{volume}{29}, \bibinfo{pages}{480--482}.
\bibitem[{Pova{\v{z}}ay et~al.(2003)Pova{\v{z}}ay, Bizheva, Hermann, Unterhuber, Sattmann, Fercher, Drexler, Schubert, Ahnelt, Mei et~al.}]{povavzay2003enhanced}
\bibinfo{author}{Pova{\v{z}}ay, B.}, \bibinfo{author}{Bizheva, K.}, \bibinfo{author}{Hermann, B.}, \bibinfo{author}{Unterhuber, A.}, \bibinfo{author}{Sattmann, H.}, \bibinfo{author}{Fercher, A.F.}, \bibinfo{author}{Drexler, W.}, \bibinfo{author}{Schubert, C.}, \bibinfo{author}{Ahnelt, P.}, \bibinfo{author}{Mei, M.}, et~al., \bibinfo{year}{2003}.
\newblock \bibinfo{title}{Enhanced visualization of choroidal vessels using ultrahigh resolution ophthalmic \uppercase{OCT} at 1050 nm}.
\newblock \bibinfo{journal}{Optics Express} \bibinfo{volume}{11}, \bibinfo{pages}{1980--1986}.
\bibitem[{Pova{\v{z}}ay et~al.(2007)Pova{\v{z}}ay, Hermann, Unterhuber, Hofer, Sattmann, Zeiler, Morgan, Falkner-Radler, Glittenberg, Blinder et~al.}]{povavzay2007three}
\bibinfo{author}{Pova{\v{z}}ay, B.}, \bibinfo{author}{Hermann, B.}, \bibinfo{author}{Unterhuber, A.}, \bibinfo{author}{Hofer, B.}, \bibinfo{author}{Sattmann, H.}, \bibinfo{author}{Zeiler, F.}, \bibinfo{author}{Morgan, J.E.}, \bibinfo{author}{Falkner-Radler, C.}, \bibinfo{author}{Glittenberg, C.}, \bibinfo{author}{Blinder, S.}, et~al., \bibinfo{year}{2007}.
\newblock \bibinfo{title}{Three-dimensional optical coherence tomography at 1050 nm versus 800 nm in retinal pathologies: enhanced performance and choroidal penetration in cataract patients}.
\newblock \bibinfo{journal}{Journal of Biomedical Optics} \bibinfo{volume}{12}, \bibinfo{pages}{041211--041211}.
\bibitem[{Schmitt and Liang(1997)}]{schmitt1997}
\bibinfo{author}{Schmitt, J.M.}, \bibinfo{author}{Liang, Z.}, \bibinfo{year}{1997}.
\newblock \bibinfo{title}{Deconvolution and enhancement of optical coherence tomograms}, in: \bibinfo{booktitle}{Coherence Domain Optical Methods in Biomedical Science and Clinical Applications}, \bibinfo{organization}{SPIE}. pp. \bibinfo{pages}{46--57}.
\bibitem[{Shi et~al.(2019)Shi, Cai, Gu, Hu, Ma, Chen and Chen}]{shi2019despecnet}
\bibinfo{author}{Shi, F.}, \bibinfo{author}{Cai, N.}, \bibinfo{author}{Gu, Y.}, \bibinfo{author}{Hu, D.}, \bibinfo{author}{Ma, Y.}, \bibinfo{author}{Chen, Y.}, \bibinfo{author}{Chen, X.}, \bibinfo{year}{2019}.
\newblock \bibinfo{title}{Despecnet: a \uppercase{CNN}-based method for speckle reduction in retinal optical coherence tomography images}.
\newblock \bibinfo{journal}{Physics in Medicine \& Biology} \bibinfo{volume}{64}, \bibinfo{pages}{175010}.
\bibitem[{Shurrab and Duwairi(2022)}]{shurrab2022self}
\bibinfo{author}{Shurrab, S.}, \bibinfo{author}{Duwairi, R.}, \bibinfo{year}{2022}.
\newblock \bibinfo{title}{Self-supervised learning methods and applications in medical imaging analysis: A survey}.
\newblock \bibinfo{journal}{PeerJ Computer Science} \bibinfo{volume}{8}, \bibinfo{pages}{e1045}.
\bibitem[{Swanson et~al.(1993)Swanson, Izatt, Hee, Huang, Lin, Schuman, Puliafito and Fujimoto}]{swanson1993vivo}
\bibinfo{author}{Swanson, E.A.}, \bibinfo{author}{Izatt, J.A.}, \bibinfo{author}{Hee, M.R.}, \bibinfo{author}{Huang, D.}, \bibinfo{author}{Lin, C.}, \bibinfo{author}{Schuman, J.}, \bibinfo{author}{Puliafito, C.}, \bibinfo{author}{Fujimoto, J.G.}, \bibinfo{year}{1993}.
\newblock \bibinfo{title}{In vivo retinal imaging by optical coherence tomography}.
\newblock \bibinfo{journal}{Optics Letters} \bibinfo{volume}{18}, \bibinfo{pages}{1864--1866}.
\bibitem[{Szkulmowski et~al.(2016)Szkulmowski, Tamborski and Wojtkowski}]{szkulmowski2016spectrometer}
\bibinfo{author}{Szkulmowski, M.}, \bibinfo{author}{Tamborski, S.}, \bibinfo{author}{Wojtkowski, M.}, \bibinfo{year}{2016}.
\newblock \bibinfo{title}{Spectrometer calibration for spectroscopic fourier domain optical coherence tomography}.
\newblock \bibinfo{journal}{Biomedical Optics Express} \bibinfo{volume}{7}, \bibinfo{pages}{5042--5054}.
\bibitem[{Taleb et~al.(2020)Taleb, Loetzsch, Danz, Severin, Gaertner, Bergner and Lippert}]{taleb20203d}
\bibinfo{author}{Taleb, A.}, \bibinfo{author}{Loetzsch, W.}, \bibinfo{author}{Danz, N.}, \bibinfo{author}{Severin, J.}, \bibinfo{author}{Gaertner, T.}, \bibinfo{author}{Bergner, B.}, \bibinfo{author}{Lippert, C.}, \bibinfo{year}{2020}.
\newblock \bibinfo{title}{3d self-supervised methods for medical imaging}.
\newblock \bibinfo{journal}{Advances in Neural Information Processing Systems} \bibinfo{volume}{33}, \bibinfo{pages}{18158--18172}.
\bibitem[{Unterhuber et~al.(2005)Unterhuber, Pova{\v{z}}ay, Hermann, Sattmann, Chavez-Pirson and Drexler}]{unterhuber2005vivo}
\bibinfo{author}{Unterhuber, A.}, \bibinfo{author}{Pova{\v{z}}ay, B.}, \bibinfo{author}{Hermann, B.}, \bibinfo{author}{Sattmann, H.}, \bibinfo{author}{Chavez-Pirson, A.}, \bibinfo{author}{Drexler, W.}, \bibinfo{year}{2005}.
\newblock \bibinfo{title}{In vivo retinal optical coherence tomography at 1040 nm-enhanced penetration into the choroid}.
\newblock \bibinfo{journal}{Optics Express} \bibinfo{volume}{13}, \bibinfo{pages}{3252--3258}.
\bibitem[{Wang et~al.(2023)Wang, Ling, Dong, Yao, Gan, Zhou and Su}]{Wang2023}
\bibinfo{author}{Wang, M.}, \bibinfo{author}{Ling, Y.}, \bibinfo{author}{Dong, Z.}, \bibinfo{author}{Yao, X.}, \bibinfo{author}{Gan, Y.}, \bibinfo{author}{Zhou, C.}, \bibinfo{author}{Su, Y.}, \bibinfo{year}{2023}.
\newblock \bibinfo{title}{\uppercase{GPU}-accelerated iterative method for \uppercase{FD-OCT} image reconstruction with an image-level cross-domain regularizer}.
\newblock \bibinfo{journal}{Optics Express} \bibinfo{volume}{31}, \bibinfo{pages}{1813--1831}.
\bibitem[{Wang et~al.(2021)Wang, Zhu, Yu, Chen, Shi, Zhou, Ma, Peng, Bao, Feng et~al.}]{wang2021semi}
\bibinfo{author}{Wang, M.}, \bibinfo{author}{Zhu, W.}, \bibinfo{author}{Yu, K.}, \bibinfo{author}{Chen, Z.}, \bibinfo{author}{Shi, F.}, \bibinfo{author}{Zhou, Y.}, \bibinfo{author}{Ma, Y.}, \bibinfo{author}{Peng, Y.}, \bibinfo{author}{Bao, D.}, \bibinfo{author}{Feng, S.}, et~al., \bibinfo{year}{2021}.
\newblock \bibinfo{title}{Semi-supervised capsule c\uppercase{GAN} for speckle noise reduction in retinal \uppercase{OCT} images}.
\newblock \bibinfo{journal}{IEEE Transactions on Medical Imaging} \bibinfo{volume}{40}, \bibinfo{pages}{1168--1183}.
\bibitem[{Wang(1999)}]{Wang1999}
\bibinfo{author}{Wang, R.K.}, \bibinfo{year}{1999}.
\newblock \bibinfo{title}{Resolution improved optical coherence-gated tomography for imaging through biological tissues}.
\newblock \bibinfo{journal}{Journal of Modern Optics} \bibinfo{volume}{46}, \bibinfo{pages}{1905--1912}.
\bibitem[{Wojtkowski et~al.(2002)Wojtkowski, Leitgeb, Kowalczyk, Bajraszewski and Fercher}]{wojtkowski2002vivo}
\bibinfo{author}{Wojtkowski, M.}, \bibinfo{author}{Leitgeb, R.}, \bibinfo{author}{Kowalczyk, A.}, \bibinfo{author}{Bajraszewski, T.}, \bibinfo{author}{Fercher, A.F.}, \bibinfo{year}{2002}.
\newblock \bibinfo{title}{In vivo human retinal imaging by fourier domain optical coherence tomography}.
\newblock \bibinfo{journal}{Journal of Biomedical Optics} \bibinfo{volume}{7}, \bibinfo{pages}{457--463}.
\bibitem[{Wojtkowski et~al.(2004)Wojtkowski, Srinivasan, Ko, Fujimoto, Kowalczyk and Duker}]{wojtkowski2004ultrahigh}
\bibinfo{author}{Wojtkowski, M.}, \bibinfo{author}{Srinivasan, V.J.}, \bibinfo{author}{Ko, T.H.}, \bibinfo{author}{Fujimoto, J.G.}, \bibinfo{author}{Kowalczyk, A.}, \bibinfo{author}{Duker, J.S.}, \bibinfo{year}{2004}.
\newblock \bibinfo{title}{Ultrahigh-resolution, high-speed, fourier domain optical coherence tomography and methods for dispersion compensation}.
\newblock \bibinfo{journal}{Optics Express} \bibinfo{volume}{12}, \bibinfo{pages}{2404--2422}.
\bibitem[{Xu et~al.(2020)Xu, Tang, Hao, Chen and Lei}]{xu2020texture}
\bibinfo{author}{Xu, M.}, \bibinfo{author}{Tang, C.}, \bibinfo{author}{Hao, F.}, \bibinfo{author}{Chen, M.}, \bibinfo{author}{Lei, Z.}, \bibinfo{year}{2020}.
\newblock \bibinfo{title}{Texture preservation and speckle reduction in poor optical coherence tomography using the convolutional neural network}.
\newblock \bibinfo{journal}{Medical Image Analysis} \bibinfo{volume}{64}, \bibinfo{pages}{101727}.
\bibitem[{Yuan et~al.(2020)Yuan, Yang, Pan and Liang}]{Yuan2020}
\bibinfo{author}{Yuan, Z.}, \bibinfo{author}{Yang, D.}, \bibinfo{author}{Pan, H.}, \bibinfo{author}{Liang, Y.}, \bibinfo{year}{2020}.
\newblock \bibinfo{title}{Axial super-resolution study for optical coherence tomography images via deep learning}.
\newblock \bibinfo{journal}{IEEE Access} \bibinfo{volume}{8}, \bibinfo{pages}{204941--204950}.
\bibitem[{Yun et~al.(2003)Yun, Tearney, Bouma, Park and de~Boer}]{yun2003high}
\bibinfo{author}{Yun, S.}, \bibinfo{author}{Tearney, G.}, \bibinfo{author}{Bouma, B.}, \bibinfo{author}{Park, B.}, \bibinfo{author}{de~Boer, J.F.}, \bibinfo{year}{2003}.
\newblock \bibinfo{title}{High-speed spectral-domain optical coherence tomography at 1.3 $\mu$m wavelength}.
\newblock \bibinfo{journal}{Optics Express} \bibinfo{volume}{11}, \bibinfo{pages}{3598--3604}.
\bibitem[{Zhang et~al.(2021)Zhang, Liu, Singh, {\c{C}}etinta{\c{s}}, Luo, Rivenson, Larin and Ozcan}]{Zhang2021}
\bibinfo{author}{Zhang, Y.}, \bibinfo{author}{Liu, T.}, \bibinfo{author}{Singh, M.}, \bibinfo{author}{{\c{C}}etinta{\c{s}}, E.}, \bibinfo{author}{Luo, Y.}, \bibinfo{author}{Rivenson, Y.}, \bibinfo{author}{Larin, K.V.}, \bibinfo{author}{Ozcan, A.}, \bibinfo{year}{2021}.
\newblock \bibinfo{title}{Neural network-based image reconstruction in swept-source optical coherence tomography using undersampled spectral data}.
\newblock \bibinfo{journal}{Light: Science \& Applications} \bibinfo{volume}{10}, \bibinfo{pages}{155}.

\end{thebibliography}

\clearpage
\onecolumn  
\captionsetup[table]{labelfont={bf},labelformat={default},labelsep=period,name={Supplementary Table.}}
\setcounter{table}{0}

\section*{Supplementary information}

\begin{table*}[h!]
\captionsetup{font=normalsize} 
    \centering
    \renewcommand{\arraystretch}{1.5}
    \setlength{\tabcolsep}{5pt}
    \caption{\label{tab5} Configuration of OCT systems used for data collection}
    \vspace{0pt}
    \captionsetup{skip=3pt} 
    \begin{tabular}{ p{1.5cm}<{\centering} p{2.5cm}<{\centering} p{3.5cm}<{\centering} p{1.5cm}<{\centering} p{1.5cm}<{\centering} p{1.5cm}<{\centering} p{2.5cm}<{\centering}}
    \hline
    Dataset No. & Sample & System $\&$ Vendor & Centra\newline wavelength\newline (nm) & Axial\newline resolution\newline ($\mu$ m) & Average\newline frames & Data source\\
    \hline
    1 & Human retina & Spectralis OCT,\newline Heidelberg & 870 & 7 & 16 & Publicly available \\
    2 & Human retina & Triton, Topcon & 1050 & 8 & 32 & Collected in\newline PUMCH\\
    3 & Human retina & DREAM OCT,\newline SVision Imaging & 1050 & 3.8 & 32 & Collected in\newline PUMCH\\
    4 & Human retina & DREAM OCT,\newline SVision Imaging & 1050 & 3.8 & 2 & Collected in\newline PUMCH \\
    5 & Swine esophagus\newline mucosa & Home-built ultrahigh-\newline resolution OCT & 830 & 2.6 & 1 & Collected in\newline BIT lab of JHU \\
    \hline
    \end{tabular}
\end{table*}

\begin{table*}[h]
\captionsetup{font=normalsize} 
    \centering
    \renewcommand{\arraystretch}{1.5}
    \setlength{\tabcolsep}{5pt}
    \caption{\label{tab6} Hyperparameters used by O-PRESS}
    \vspace{0pt}
    \captionsetup{skip=3pt} 
    \begin{tabular}{ p{3cm}<{\centering} p{2cm}<{\centering} p{2cm}<{\centering} p{2.2cm}<{\centering} p{2cm}<{\centering} p{2cm}<{\centering} }
    \hline
     & Model & Learning rate $\eta$ & Weight of MC loss $\lambda_1$ & Weight of EI loss $\lambda_2$ & Weight of FS loss $\lambda_3$ \\
    \hline
    know PSF & $Model_{ei}$ & $5e^{-4}$ & 1 & 1 & None \\
    Unknown PSF & $Model_s$ & $5e^{-4}$ & 1 & 10 & 10 \\
    Unknown PSF & $Model_r$ & $1e^{-3}$ & 1 & 10 & 10 \\
    \hline
    \end{tabular}
\end{table*}

\captionsetup[figure]{labelfont={bf},labelformat={default},labelsep=period,name={Supplementary Fig.}}
\setcounter{figure}{0}
\begin{figure*}[h]
    \centering
    \includegraphics[width=\textwidth]{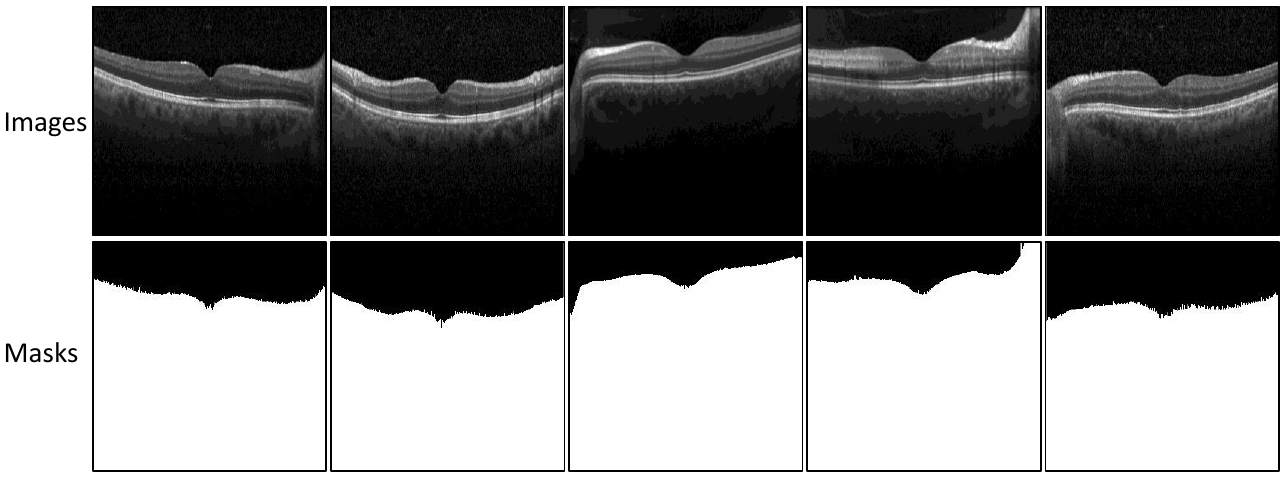}
\caption{
\begin{small}
    \textbf{Illustrated OCT images alongside their corresponding generated masks.}
\end{small}
 } 
    \label{fig:mask}
\end{figure*}

\begin{figure*}[h]
    \centering
    \includegraphics[width=\textwidth]{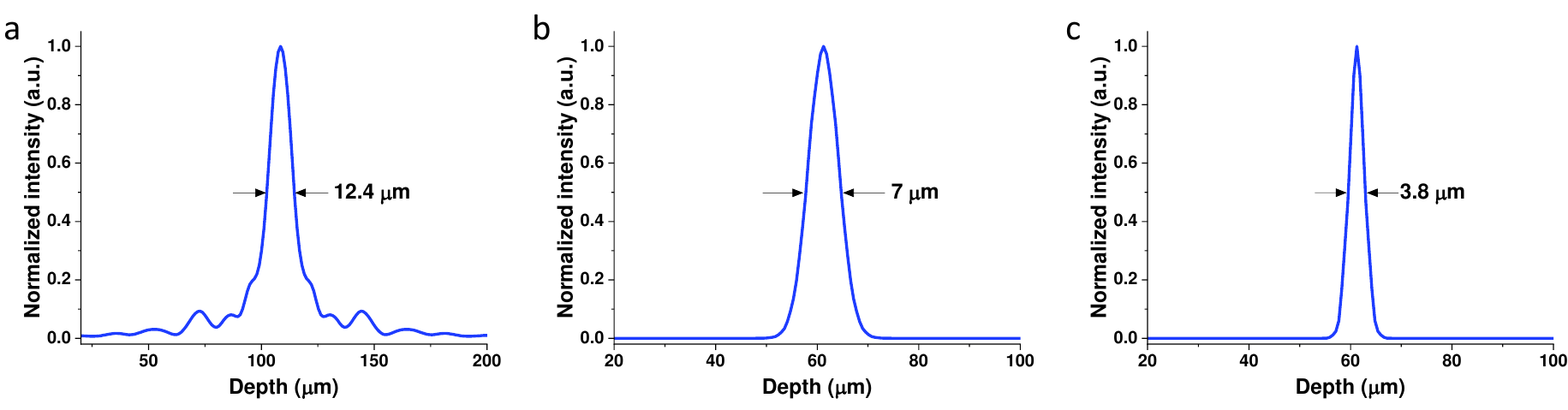}
\caption{
\begin{small}
\textbf{Point spread functions (PSFs) utilized in training. a}, A real-world PSF employed for simulating low-resolution OCT images.\textbf{ b}, Estimated PSF utilized for single-time inference. \textbf{c}, Estimated PSF employed for recurrent inference.
\end{small} }
    \label{fig:psf}
\end{figure*}

\begin{figure*}[h]
    \centering
    \includegraphics[scale=0.5]{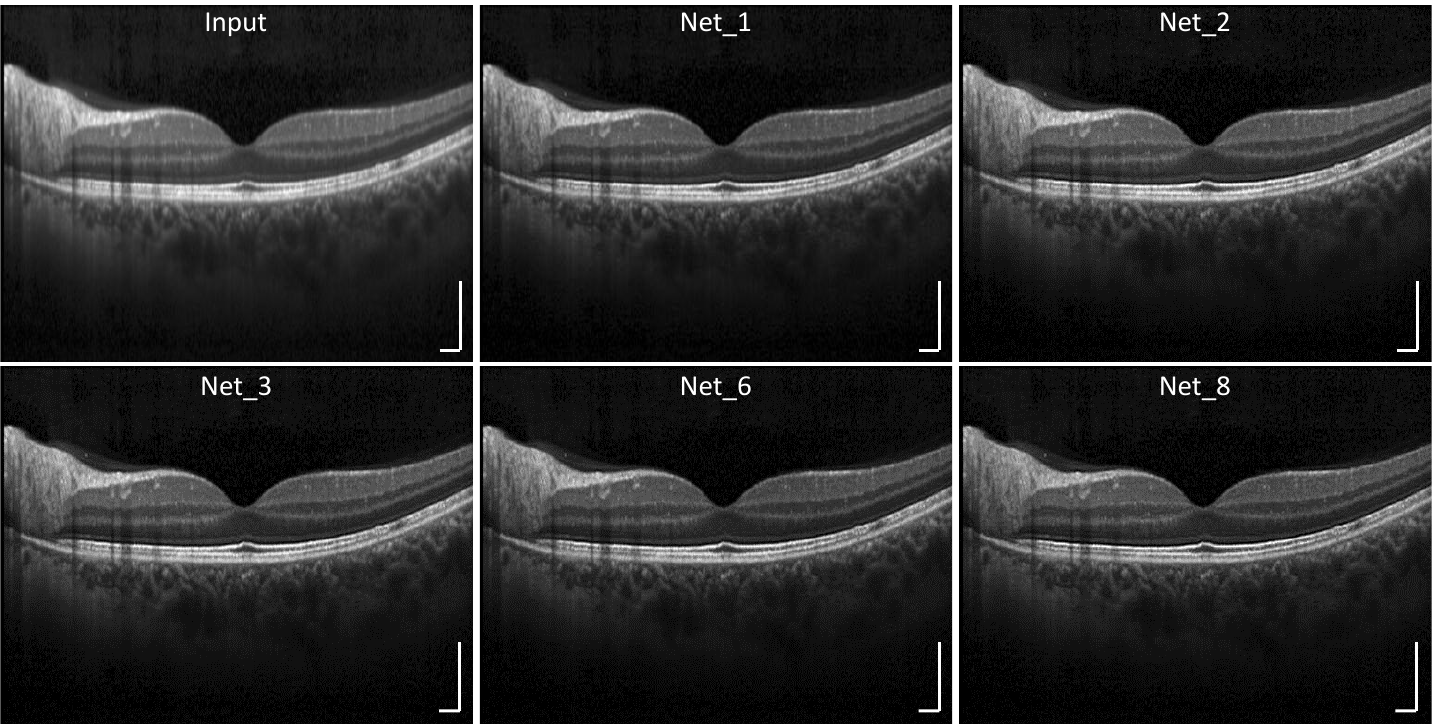}
\caption{
\begin{small}
Performance of recurrent inference. Net\_1 = $\Phi$(Input), Net\_2 = $\Phi$(0.3 Input + 0.7 Net\_1), Net\_3 = $\Phi$(0.3 Input + 0.7 Net\_2), Net\_6 = $\Phi$(0.3 Input + 0.7 Net\_5), Net\_8 = $\Phi$(0.3 Input + 0.7 Net\_7).  
Scale bar: 0.5 mm.
\end{small}}
    \label{fig:recurrent}
\end{figure*}

\begin{figure*}[h]
    \centering
    \includegraphics[width=\textwidth]{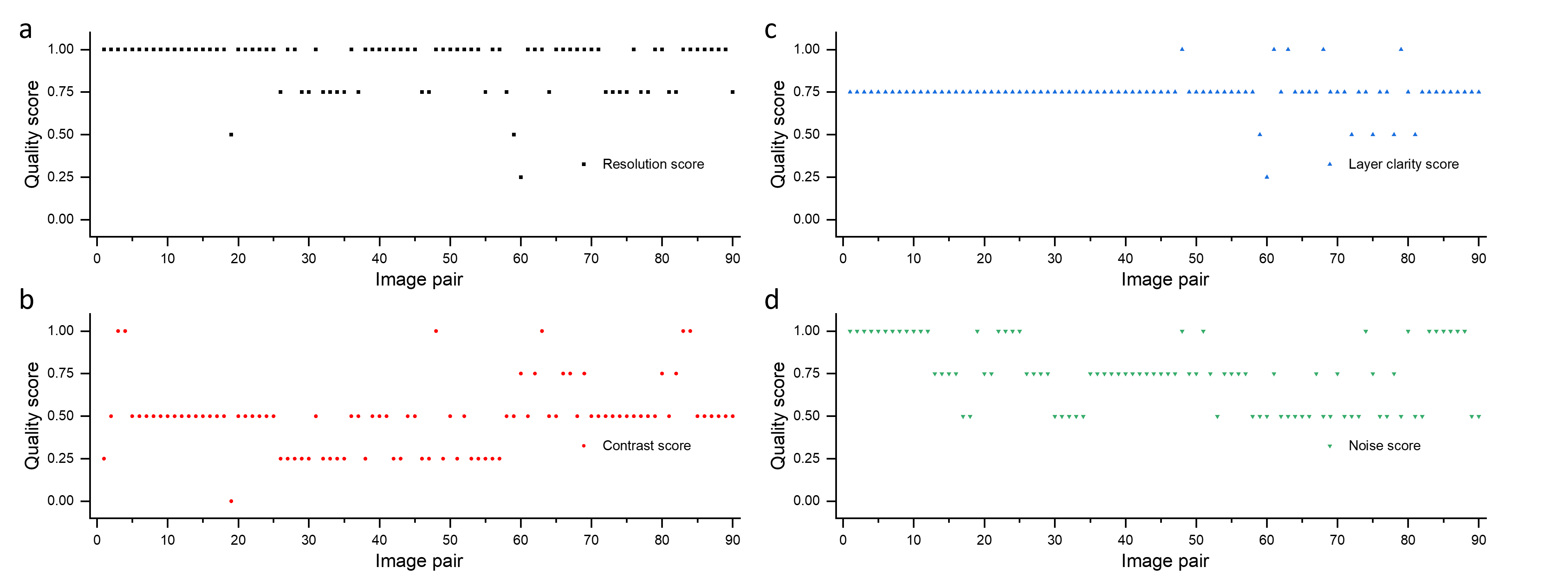}
\caption{
\begin{small}
    \textbf{Outcome of the perceptual study by optical experts. a}, Detailed numeric values for the resolution evaluation criterion. \textbf{b}, Detailed numeric values for the contrast evaluation criterion. \textbf{c}, Detailed numeric values for the layer clarity evaluation criterion. \textbf{d}, Detailed numeric values for the noise level evaluation criterion.
\end{small}
}
    \label{fig:pecp_optics}
\end{figure*}

\begin{figure*}[h]
    \centering
    \includegraphics[width=\textwidth]{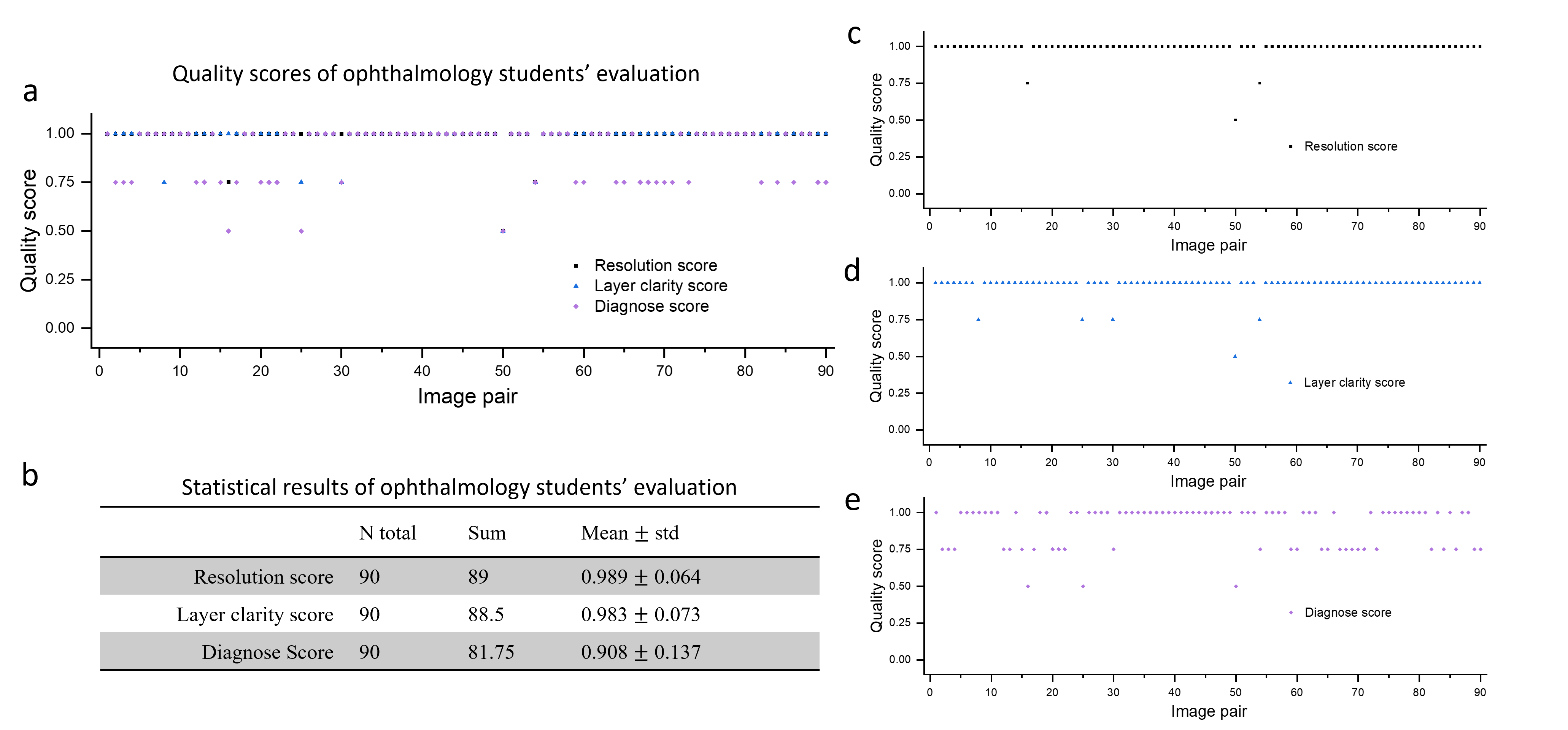}
\caption{
\begin{small}
    \textbf{Outcome of the perceptual study by ophthalmology students. a}, Quality scores for all evaluation criteria. \textbf{b}, Summary of the statistical analysis.  \textbf{c}, Detailed numeric values for the resolution evaluation criterion. \textbf{d}, Detailed numeric values for the layer clarity evaluation criterion. \textbf{e}, Detailed numeric values for the diagnose evaluation criterion.
\end{small}
} 
    \label{fig:pecp_stud}
\end{figure*}

\begin{figure*}[h]
    \centering
    \includegraphics[width=\textwidth]{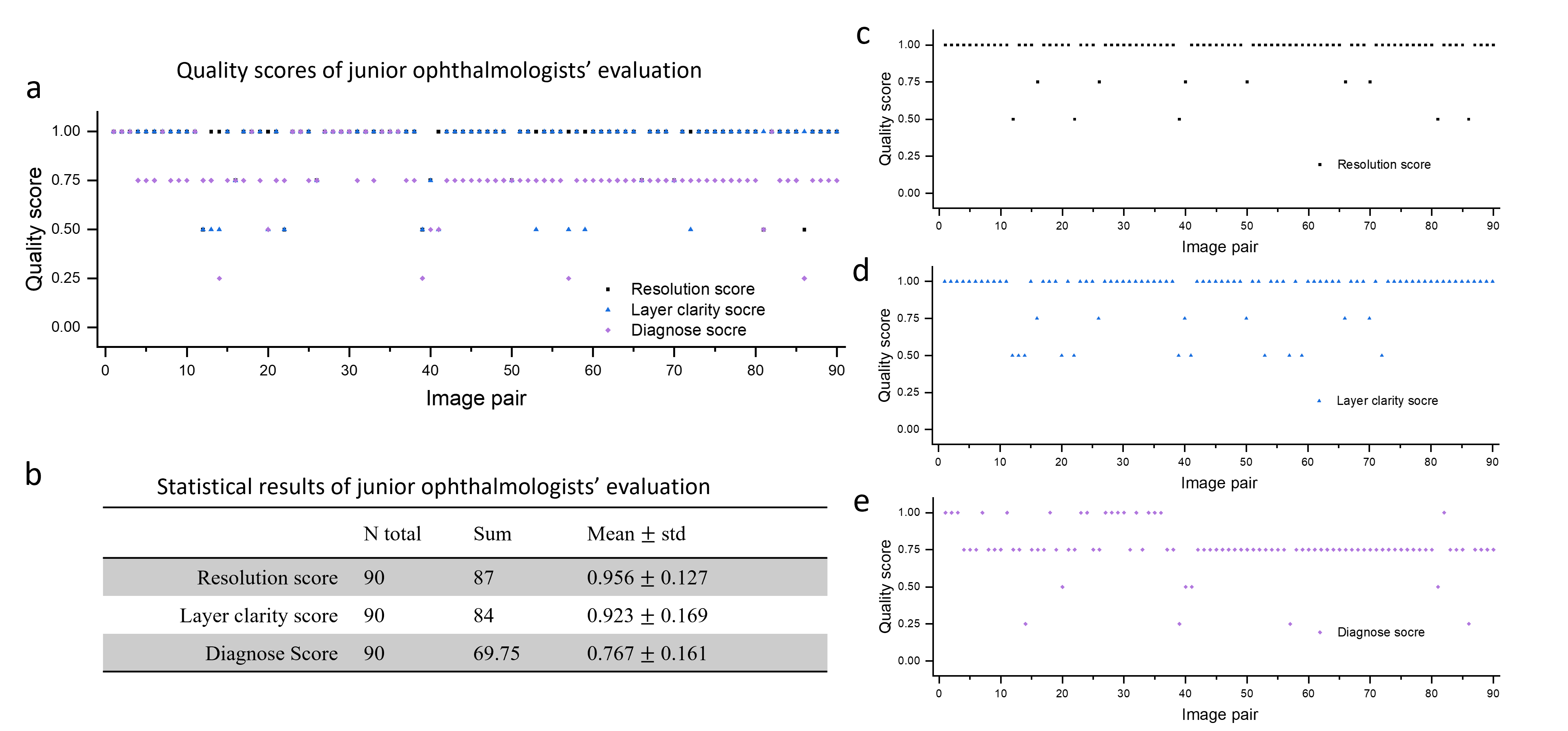}
\caption{
\begin{small}
    Outcome of the perceptual study by ophthalmology residents . \textbf{a}, Quality scores for all evaluation criteria. \textbf{b}, Summary of the statistical analysis. \textbf{c}, Detailed numeric values for the resolution evaluation criterion. \textbf{d}, Detailed numeric values for the layer clarity evaluation criterion. \textbf{e}, Detailed numeric values for the diagnose evaluation criterion.
\end{small}} 
    \label{fig:pecp_junior}
\end{figure*}

\begin{figure*}[h]
    \centering
    \includegraphics[width=\textwidth]{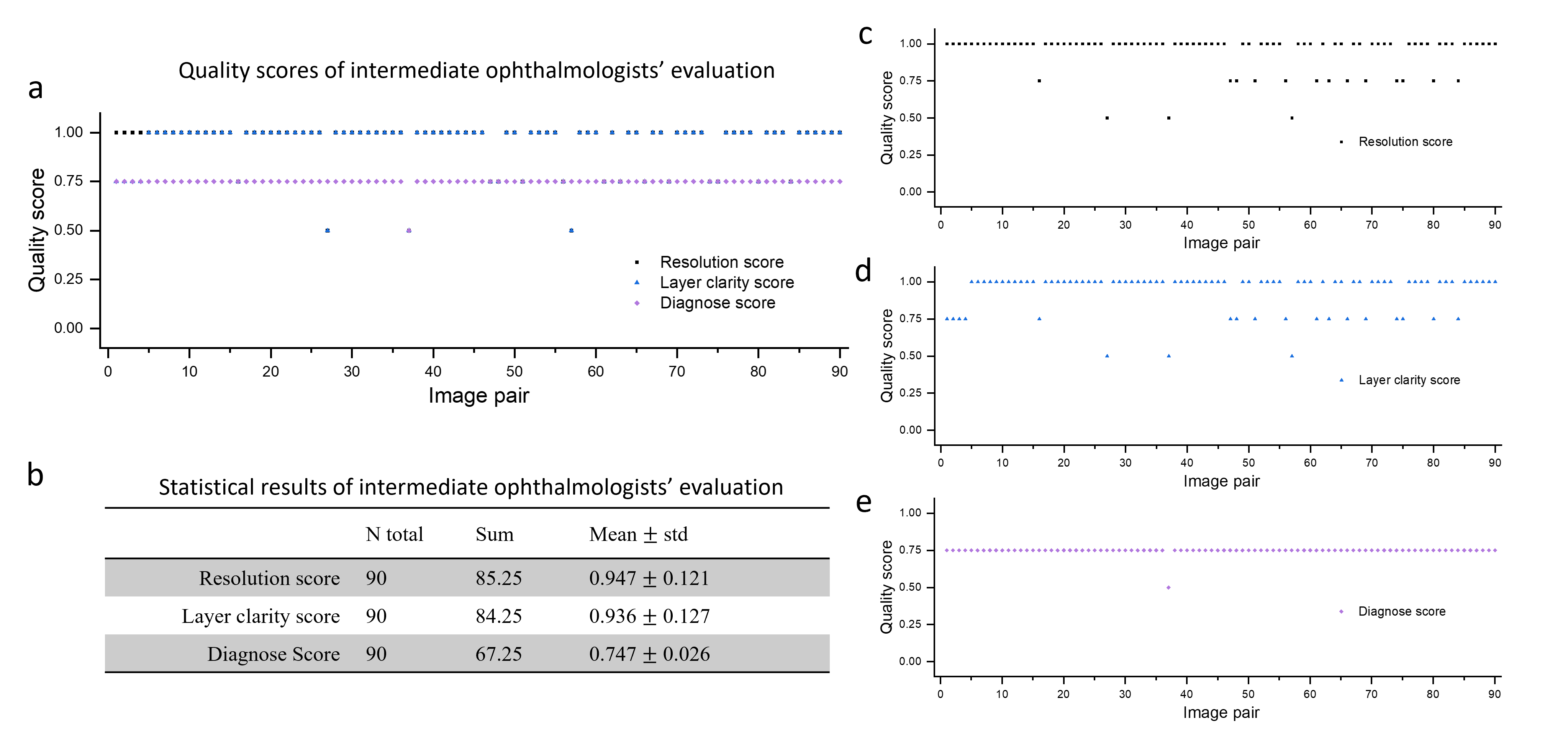}
\caption{
\begin{small}
    \textbf{Outcome of the perceptual study by ophthalmology fellows. a}, Quality scores for all evaluation criteria. \textbf{b}, Summary of the statistical analysis. \textbf{c}, Detailed numeric values for the resolution evaluation criterion. \textbf{d}, Detailed numeric values for the layer clarity evaluation criterion. \textbf{e}, Detailed numeric values for the diagnose evaluation criterion.
\end{small}
} 
    \label{fig:pecp_intermediate}
\end{figure*}

\begin{figure*}[h]
    \centering
    \includegraphics[width=\textwidth]{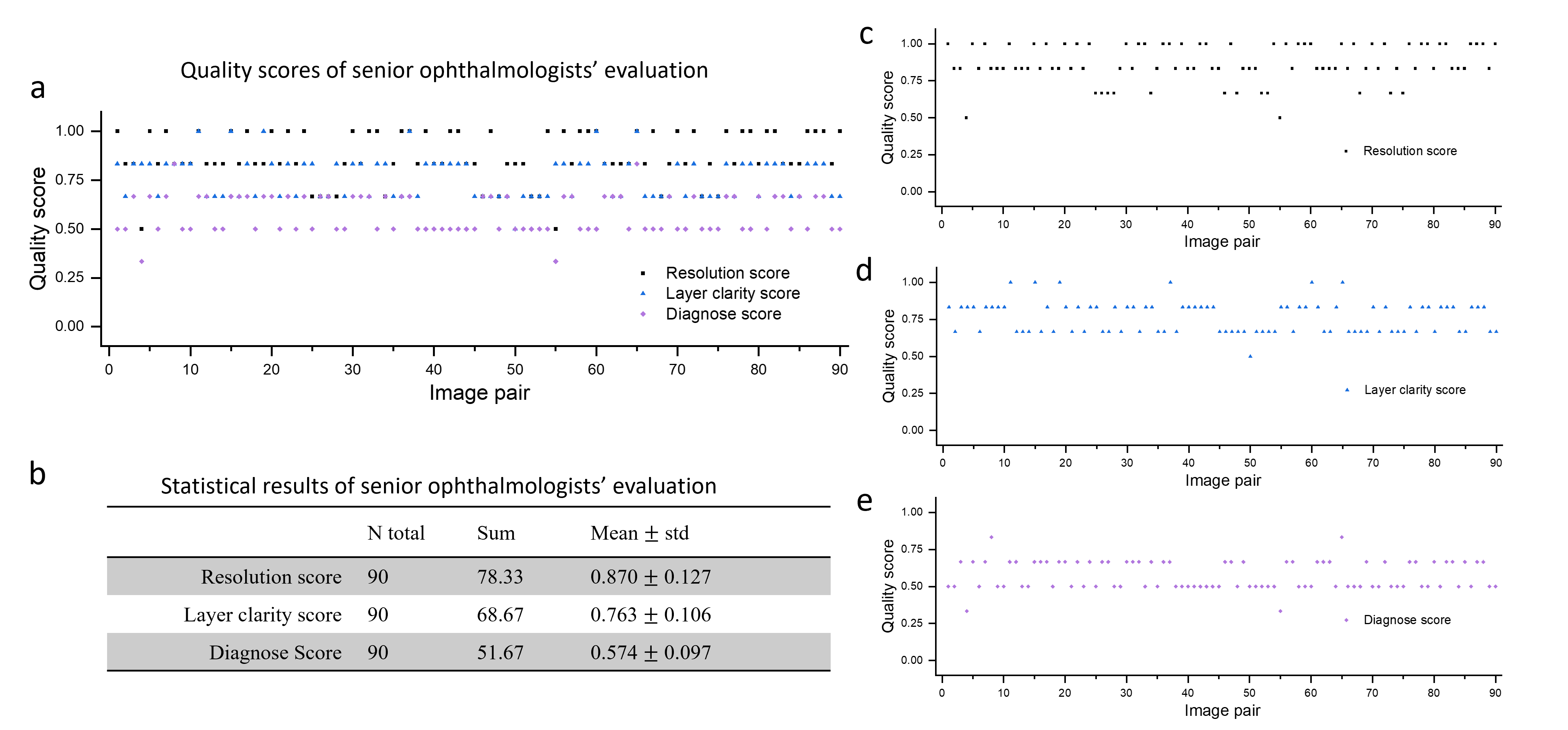}
\caption{
\begin{small}
    \textbf{Outcome of the perceptual study by senior ophthalmologists. a}, Quality scores for all evaluation criteria. \textbf{b}, Summary of the statistical analysis. \textbf{c}, Detailed numeric values for the resolution evaluation criterion. \textbf{d}, Detailed numeric values for the layer clarity evaluation criterion. \textbf{e}, Detailed numeric values for the diagnose evaluation criterion.
\end{small}
 } 
    \label{fig:pecp_senior}
\end{figure*}

\begin{figure*}[h]
    \centering
    \includegraphics[scale=0.6]{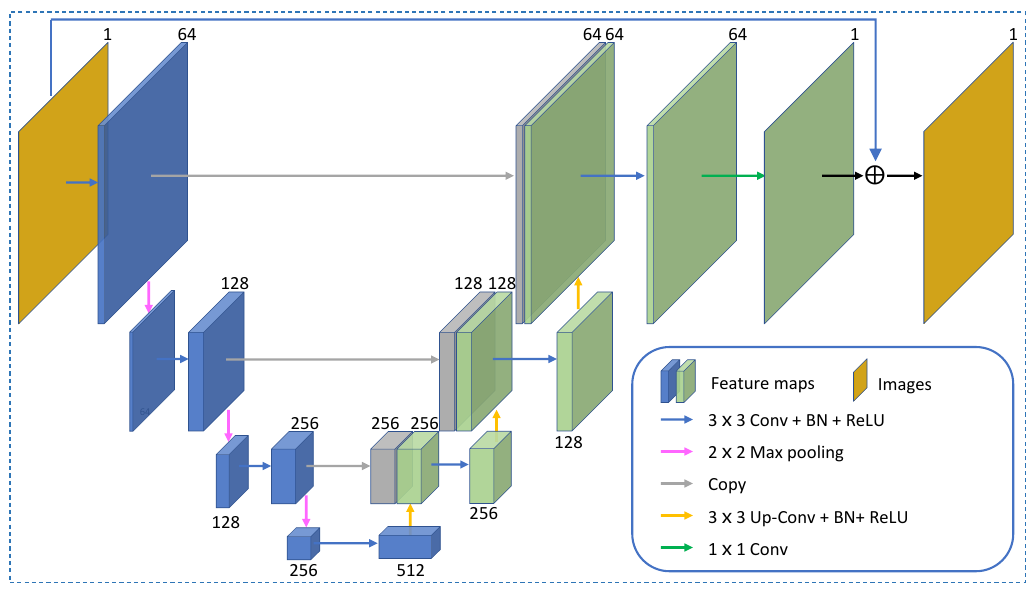}
\caption{
\begin{small}
    \textbf{The network architecture used in the proposed approach.} Channel numbers of the Res-UNet are indicated above or below the feature maps and images.
\end{small}
} 
    \label{fig:ResUnet}
\end{figure*}

\end{document}